\documentclass[twocolumn,tighten]{aastex631}
\usepackage{graphicx}	% Including figure files
\usepackage{amsmath}	% Advanced maths commands
\usepackage{amssymb}	% Extra maths symbols
\usepackage{bm}		% Bold maths symbols, including upright Greek
\usepackage{rotating}
\usepackage{longtable}

\def\lea{\mathrel{<\kern-1.0em\lower0.9ex\hbox{$\sim$}}}
\def\gea{\mathrel{>\kern-1.0em\lower0.9ex\hbox{$\sim$}}}

\newcommand{\UWyoming}{\affiliation{Department of Physics and Astronomy, University of Wyoming, Laramie, WY 82071, USA}}
\newcommand{\STScI}{\affiliation{Space Telescope Science Institute, 3700 San Martin Drive, Baltimore, MD 21218, USA}}

\newcommand{\UToledo}{\affiliation{Ritter Astrophysical Research Center, University of Toledo, Toledo, OH 43606, USA}}
\newcommand{\JHU}{\affiliation{Department of Physics and Astronomy, The Johns Hopkins University, Baltimore, MD 21218 USA}}

\newcommand{\OSU}{\affiliation{Department of Astronomy, The Ohio State University, 140 West 18th Ave., Columbus, OH 43210, USA}}
\newcommand{\OSUphys}
{\affiliation{Department of Physics, The Ohio State University, Columbus, Ohio 43210, USA}}
\newcommand{\OSUcosmo}
{\affiliation{Center for Cosmology \& Astro-Particle Physics, The Ohio State University, Columbus, Ohio 43210, USA}}
\newcommand{\UMichigan}{\affiliation{Department of Astronomy, University of Michigan, Ann Arbor, MI 48109, USA}}
\newcommand{\MPIA}{\affiliation{Max Planck Institut f\"ur Astronomie, K\"onigstuhl 17, 69117 Heidelberg, Germany}}

\newcommand{\UCSD}{\affiliation{Center for Astrophysics \& Space Sciences, Department of Physics,  University of California San Diego, 9500 Gilman Drive, La Jolla, CA 92093, USA}}

\newcommand{\Bonn}{\affiliation{Argelander-Institut für Astronomie, Universität Bonn, Auf dem Hügel 71, 53121, Bonn, Germany}}
\newcommand{\ANU}{\affiliation{Research School of Astronomy and Astrophysics, Australian National University, Canberra, ACT 2611, Australia}}

\newcommand{\ITA}{\affiliation{Institut f\"{u}r Theoretische Astrophysik, Zentrum f\"{u}r Astronomie der Universit\"{a}t Heidelberg,\\ Albert-Ueberle-Strasse 2, 69120 Heidelberg, Germany}}
\newcommand{\COOL}{\affiliation{Cosmic Origins Of Life (COOL) Research DAO, \href{https://coolresearch.io}{https://coolresearch.io}}}

\newcommand{\UArizona}{\affiliation{Steward Observatory, University of Arizona, 933 N Cherry Ave,Tucson, AZ 85721, USA}}

\newcommand{\LaPlata}{\affiliation{Instituto de Astrofisica de La Plata, CONICET–UNLP,
 Paseo del Bosque S/N, B1900FWA La Plata, Argentina }}
\newcommand{\sorbonne}{\affiliation{Sorbonne {Universit\'e}, LERMA, Observatoire de Paris, PSL university, CNRS, F-75014, Paris, France}}

\newcommand{\UNAM}{\affiliation{Instituto de Astronom\'ia, Universidad Nacional Aut\'onoma de M\'exico, Unidad Acad\'emica en Ensenada, Km 103 Carr. Tijuana−Ensenada, Ensenada, B.C.,
C.P. 22860, M\'exico}}
\newcommand{\Whitman}{\affil{Whitman College, 345 Boyer Avenue, Walla Walla, WA 99362, USA}}
\newcommand{\ARC}{\affil{ARC Centre of Excellence for All Sky Astrophysics in 3 Dimensions (ASTRO 3D), Australia}}   
\newcommand{\stsciesa}{\affil{AURA for the European Space Agency (ESA), Space Telescope Science Institute, 3700 San Martin Drive, Baltimore, MD 21218, USA}}
\newcommand{\oxford}{\affil{Department of Physics, University of Oxford, Keble Road, Oxford OX1 3RH, UK}}
\newcommand{\Carnegie}{\affil{The Observatories of the Carnegie Institution for Science, 813 Santa Barbara Street, Pasadena, CA 91101, USA}}
\newcommand{\Virginia}{\affil{University of Virginia, Charlottesville, VA, USA }}
\newcommand{\Rechen}{\affil{Astronomisches Rechen-Institut, Zentrum für Astronomie der Universität Heidelberg, Mönchhofstraße 12-14, D-69120 Heidelberg, Germany}}
\newcommand{\UniCA}{\affil{Université Côte d'Azur, Observatoire de la Côte d'Azur, CNRS, Laboratoire Lagrange, 06000, Nice, France}}
\newcommand{\Alberta}{\affil{Dept. of Physics, University of Alberta, 4-183 CCIS, Edmonton, Alberta, T6G 2E1, Canada}}
\newcommand{\Gent}{\affil{Sterrenkundig Observatorium, Universiteit Gent, Krijgslaan 281 S9, B-9000 Gent, Belgium}}
\shorttitle{Empirical SED Templates from HST and JWST}
\shortauthors{Whitmore et al.}

\DeclareUnicodeCharacter{2212}{-}
\begin{document}

% No PAH or Infrared Dust Emission in Star Clusters Older than Five Myr or 
\title{Empirical SED Templates for Star Clusters Observed with HST and JWST: No Strong PAH or IR Dust Emission after Five Myr}

% Author List

\correspondingauthor{Bradley~C.~Whitmore}
\email{whitmore@stsci.edu}
\author[0000-0002-3784-7032]{Bradley~C.~Whitmore}
\STScI
\author[0000-0003-0085-4623]{Rupali~Chandar}
\UToledo
\author[0000-0003-0946-6176]{Janice~C.~Lee}
\STScI
\UArizona
\author[0000-0001-7448-1749]{Kiana~F.~Henny}
\UWyoming
\author[0000-0002-0579-6613]{M. Jimena Rodríguez}
\STScI
\LaPlata
\author[0000-0003-4974-3481]{Dalya Baron}
\Carnegie
\author[0000-0003-0166-9745]{F. Bigiel}
\Bonn
\author[0000-0003-0946-6176]{M\'ed\'eric~Boquien}
\UniCA
\author[0000-0002-5635-5180]{M\'elanie Chevance}
\ITA
\COOL
\author[0000-0000-0000-0000]{Ryan~Chown}
\OSU
\author[0000-0002-5782-9093]{Daniel~A.~Dale}
\UWyoming
\author{Matthew~Floyd}
\UToledo
\author[0000-0002-3247-5321]{Kathryn~Grasha}
\ARC 
\ANU
\author[0000-0001-6708-1317]{Simon C.~O. Glover}
\ITA
\author[0000-0001-9852-9954]{Oleg Gnedin}
\UMichigan
\author[0000-0002-8806-6308]{Hamid Hassani}
\Alberta
\author[0000-0002-4663-6827]{Remy Indebetouw}
\Virginia
\author[0000-0002-5187-1725]{Anand Utsav Kapoor}
\Gent
\author[0000-0003-3917-6460]{Kirsten~L.~Larson}
\stsciesa
\author[0000-0002-2545-1700]{Adam~K.~Leroy}
\OSU
\author[0000-0001-6038-9511]{Daniel Maschmann}
\UArizona
\sorbonne
\author[0000-0003-2707-4678]{Fabian~Scheuermann}
\Rechen
\author[0000-0002-9183-8102]{Jessica Sutter}
\UCSD
\Whitman
\author[0000-0002-3933-7677]{Eva~Schinnerer}
\MPIA
\author[0000-0002-4781-7291]{Sumit K. Sarbadhicary}
\OSU
\OSUphys
\OSUcosmo
\author[0000-0002-8528-7340]{David~A.~Thilker}
\JHU
\author[0000-0002-0012-2142]{Thomas~G.~Williams}
\oxford 
\author[0000-0001-8289-3428]{Aida Wofford}
\UNAM
\UCSD

\begin{abstract} 
JWST observations, when combined with 
HST data, promise to improve age estimates of star clusters in nearby spiral galaxies.
However, feedback from young cluster stars pushes out the natal gas and dust, 
% creating complex bubbles and shells which make 
making cluster formation and evolution a challenge to model. 
Here, we use JWST$+$ HST observations of the nearby spiral galaxy NGC~628 to produce 
spectral energy distribution (SED) templates of compact star clusters 
spanning 
275 nm through 21 $\mu$m. These preliminary SEDs capture the cluster stars and associated gas and dust within radii of $\approx0.12^{\prime\prime}$ to $0.67^{\prime\prime}$ (corresponding to $\approx6$ to $33$~pc at the distance of NGC~628).
%, and can be used to age-date clusters. 
One important finding is that the SEDs of 1, 2, 3, and 4~Myr clusters can be differentiated in the infrared.  Another is that in 80-90\% of the cases we study, the PAH and H$_{\alpha}$ emission track one another, with the dust responsible for the 3.3 $\micron$ PAH emission largely removed  by  4 Myr, consistent with pre-supernova stellar feedback acting quickly on the surrounding gas and dust.  Nearly-embedded cluster candidates have infrared SEDs which are quite similar to optically visible 1 to 3~Myr clusters. In nearly all cases we find there is a young star cluster within a few tenths of an arcsec (10 - 30 pc) of the nearly embedded cluster, suggesting  the formation of the cluster was triggered by its presence. The resulting age estimates from the empirical templates are compatible both with dynamical estimates based on CO superbubble expansion velocities, and the TODDLERS models which track spherical evolution of homogeneous gas clouds around young stellar clusters. 

\end{abstract}

%\keywords{ galaxies: star formation -- galaxies: star clusters: general }

\section{Introduction and Motivation}
\label{sec:intro}

%para 1: 
Age-dating star clusters in spirals and other actively star-forming galaxies provides direct physical insight into the formation and evolution of the clusters, as well as the dynamical structures, evolution, and star-formation history of the galaxy itself.  The very youngest clusters, those younger than $\approx3$~Myr, also provide important 
% insight 
answers to questions about the timescales for interaction between stars and the interstellar medium.  Recently formed star clusters help to constrain the time it takes for giant molecular clouds to form young clusters and the time it takes for stellar feedback to halt star formation (which conserves gas for future star formation) and disperse the parent cloud \citep{kawamura09, whitmore14, hollyhead15, grasha18, Matthews18, grasha19, Chevance20, turner21,  messa21, kim22, sun24}.
However, cluster age-dating has proven to be challenging using near-ultraviolet through optical broadband photometry alone 
(e.g., \citealt{worthey94, whitmore02, anders04a,   wofford16, adamo17, whitmore20, whitmore23b}).  

% para 2:
The Physics at High Angular Resolution in Nearby GalaxieS with the Hubble Space Telescope program \citep[PHANGS-HST][G0-15654]{lee22} has used photometric measurements in five broad-band filters (NUV, U, B, V, and I) to determine the best fit age and reddening for clusters in 38 galaxies by fitting to predictions from the \citet{bruzual03} evolutionary models \citep{turner21}.  They found that without additional constraints and information, such as H$_{\alpha}$ or CO, it is challenging to separate reddened young clusters from older clusters with little reddening \citep{hannon22, whitmore23b, floyd24} and Thilker et al. 2025 (submitted),
% (Whitmore23, Thilker24), 
and there is little ability to differentiate the ages of the youngest $<5$~Myr clusters.

% para 3:
Infrared measurements of star clusters have long-promised to break the well-known age-reddening degeneracy when combined with optical data, since starlight is significantly less extinguished in the infrared. 
Observations with JWST can capture emission from starlight (at shorter infrared wavelengths), warm ionized gas (from hydrogen recombination lines), warm dust and PAH emission (Polycyclic Aromatic Hydrocarbons). Including all these components in a consistent manner is our primary challenge.

PAHs are grains that emit prominently in the mid-infrared \citep{leger84, puget89, desert90, draine03, li20, draine21}.  
The PHANGS-JWST survey (GO-2107; PI: J. Lee - \citealt{Lee23, williams24}) includes the F335M, F770W, and F1130W filters to study PAH emission on the physical scales of star clusters.
PAHs range in sizes from $\sim$3~\AA\ to $\sim$100~\AA\, with the smallest grains emitting predominantly at 3.3 $\mu$m and being particularly sensitive to ultraviolet radiation (e.g., \citealt{draine21}; \citealt{hensley23}). 
There appear to be two primary morphological signatures of PAH and IR dust emission in spiral galaxies. These have been termed ``meatballs'' (bright, roughly spherical, generally in active star-forming regions)  and ``swiss cheese'' (faint, diffuse, generally in dust lanes)   (\citealt{sandstrom23a}; \citealt{leroy23}; \citealt{belfiore22}; \citealt{pathak24};  \citealt{schinnerer24}). In the current paper we   focus on the ``meatball'' morphology, which is generally associated with HII regions and young clusters.

The spatial resolution of a telescope varies linearly with wavelength. For this reason, early Infra-Red (IR) studies had limited resolution, ranging from $\approx$ 5$^{\prime}$ for IRAS \citep{neugebauer84}, to $\approx$ 2$^{\prime\prime}$ for SPITZER \citep{werner04}, to $\approx$ 5$^{\prime\prime}$ for Herschel \citep{pilbratt10}. One of the results of this limited resolution was a focus on developing SED models that were appropriate for large portions of nearby galaxies (or the entire galaxy) rather than  individual star clusters.  Only with the 20-fold improvement in spatial resolution offered by JWST are we now able to focus on individual star clusters in nearby spiral galaxies, and the individual parts of HII region complexes.

Most current IR SED models, such as the CIGALE  \citep[Code Investigation Galaxy Emission]{boquien19} implementation of the Draine models \citep{draine07}  
are static (rather than dynamical) models which 
assume ``energy balance" between the  radiation field (primarily in the UV)
and emission from dust. While these models have been very successful for entire galaxies and kpc-scale regions in nearby galaxies that contain an integrated population containing both young, massive, UV-bright stars and the gas and dust they energize, they may not be appropriate for the study of individual star clusters on parsec scales in nearby spiral galaxies that drive outflows and typically clear the gas and dust out of the measurement aperture used to study star clusters on timescales of 2 – 5 My. 
Hence, the assumption of energy balance 
may  be violated at scales which resolve star clusters.

% para 5:
The production of empirical spectral energy distributions (SEDs) of star clusters with independently known ages, and which cover the ultraviolet, optical, and infrared portions of the spectrum, is an important step toward accurately age-dating cluster populations in nearby galaxies.
% and for testing model predictions.
The focus of this paper is to create a new set of SEDs for star clusters that span from 275 nm through 21 $\mu$m using 14 band photometry from the PHANGS-HST and JWST programs.
%the near-ultraviolet through the near- and mid-infrared parts of the spectrum. 
These can be used to 
 age-date optically visible, partially obscured, and nearly embedded star clusters. They can also be used 
 % including in dusty regions of star-forming galaxies,
 to help validate hydrodynamic simulations of star cluster formation (e.g., STARFORGE --   \citealt{Grudic21, Grudic22}), and  
  SED-generating simulations such as TODDLERS \citep{kapoor23} or WARPFIELD-EMP \citep{Pellegrini2020}. 
  %which is designed to model the dynamic situation of outflows producing bubbles.  
  The current paper is a pilot study which focuses on 40 star clusters in the well-studied ``Phantom galaxy" NGC~628.
  %uses HST and JWST observations of the Phantom galaxy NGC~628. 
  The project will be extended in the future to include additional galaxies  and to include larger cluster samples.

The rest of this paper is organized as follows. In Section~2 we  summarize the JWST$+$HST observations,  cluster catalogs and photometry, and the method of estimating cluster ages that is used in this work. 
In Section~3 we present our main results, which includes the construction of new empirical template spectral energy distributions for star clusters. 

In Section~4 we compare our empirical SEDs with predictions  from the
the TODDLERS (Time evolution of Observables
including Dust Diagnostics and Line Emission from Regions containing young Stars) SED library \citep{kapoor23} model suite, and 
discuss previous observational work in the context of our new cluster templates.
We summarize our main results and discuss future work in Section~5.

%\section{Observations and Current SED Modelling Techniques}
\section{Data and Cluster Training Sets}
\label{data}

\begin{figure*}
\begin{center}
\includegraphics[width =7in , angle= 0]{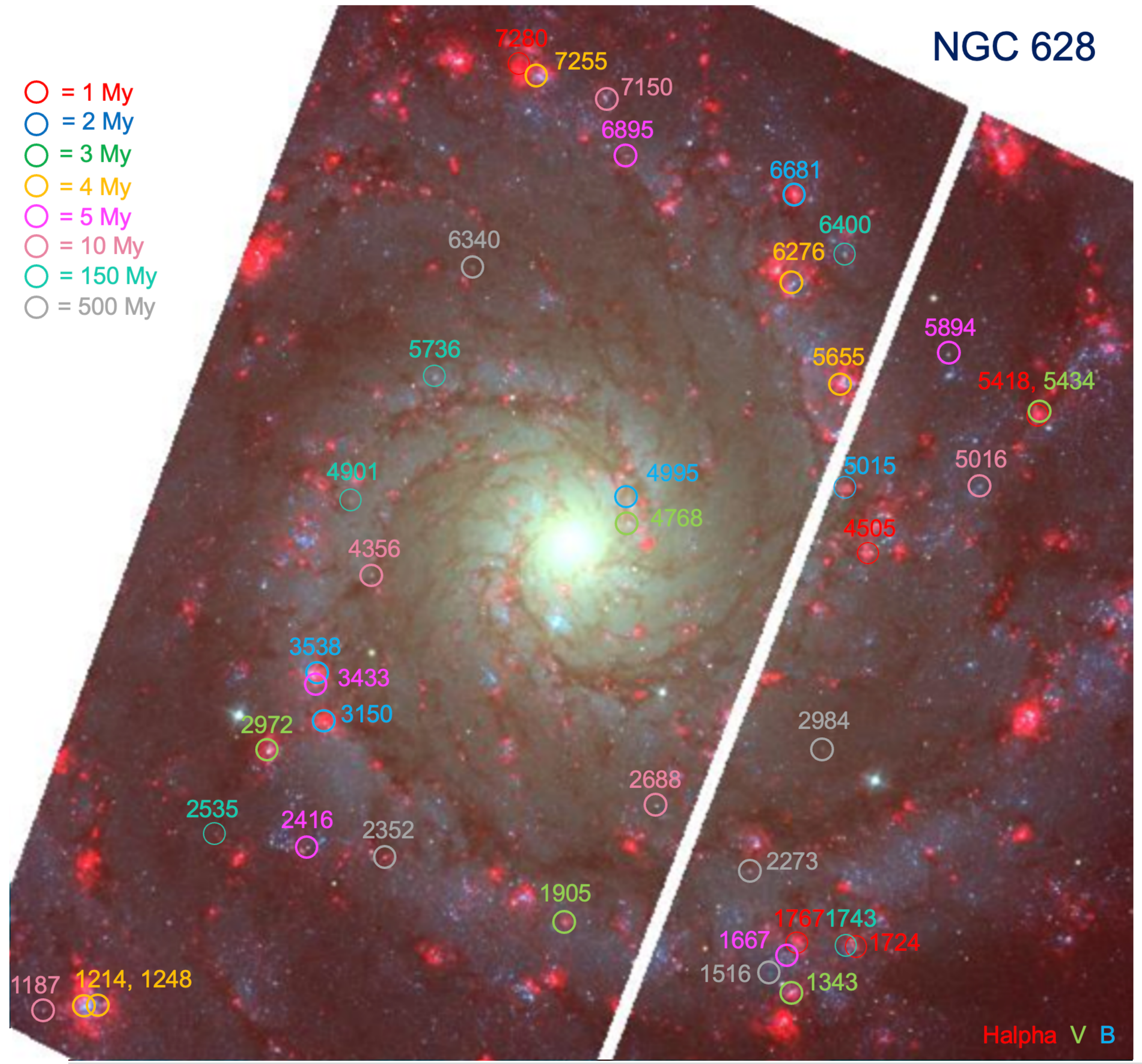}
\
\end{center}
\caption{H${\alpha}$,  V,  B HST image of a portion of NGC 628 % used to extract 
showing the locations of star clusters in our training set of Class 1 and 2 PHANGS-HST clusters for building our empirical SED templates. The color-coded  adopted ages, derived as explained in Section \ref{sec:non-degenerate}, are marked by circles, with the ID numbers from the human-classified compact cluster catalog included. Close up snapshots of the clusters are included in Figures \ref{fig:1_4_templates} and \ref{fig:5_500_templates}. 
}
\label{fig:image_and_ages_halpha}
\end{figure*}

\begin{figure*}
\begin{center}
\includegraphics[width =7in , angle= 0]{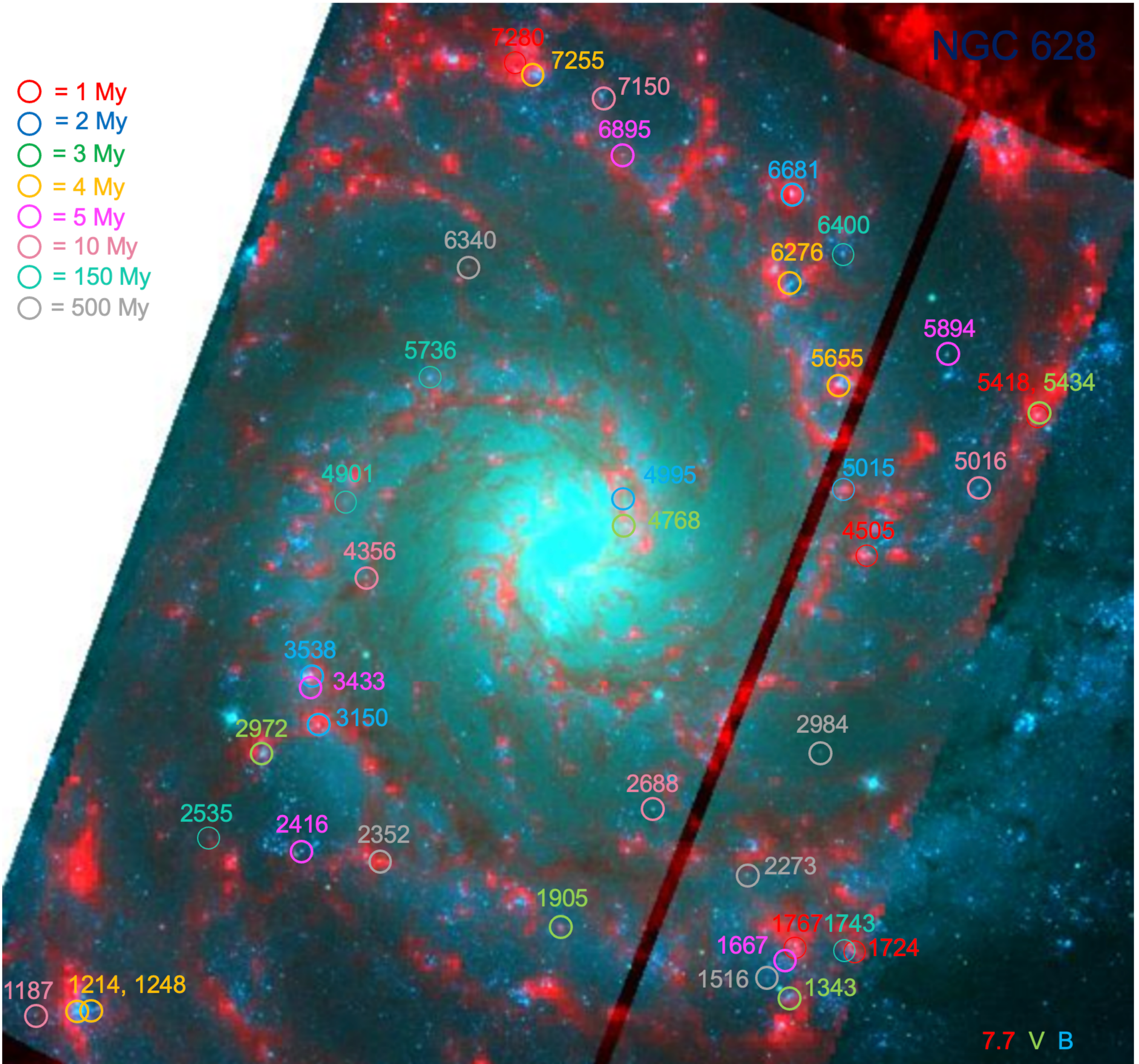}
\
\end{center}
\caption{7.7 $\micron$ (JWST), V,  B image of the same region as shown in Figure \ref{fig:image_and_ages_halpha}, with the same labeling. Note that the bottom right portion of the image was not covered by the JWST observations, limiting the region of NGC 628 that could be used to define the training set. }
\label{fig:image_and_ages_770}
\end{figure*}

\subsection{HST and JWST Observations and Cluster Catalogs}
%Photometry}
\label{sec:observations}

In this work we use photometry of clusters in NGC~628 
in 14 filters which cover wavelengths from 0.275 - 21 $\mu$m, performed on HST$+$JWST images.  NGC~628, also known as the Phantom galaxy, is a nearly face-on, grand-design spiral galaxy at a distance of $9.84\pm0.63$ Mpc  \citep{gagandeep21}. It has been the focus of several studies made by the PHANGS collaboration, as well as the LEGUS \citep{calzetti15} and FEAST \citep{gregg24}, Adamo et al. (2025 in preparation) projects.   Figure~1 shows a B, V, H$_{\alpha}$ color image of NGC~628, highlighting the HII regions strung along the spiral arms in red (H$_{\alpha}$).  Figure~2 shows a  B, V, 7.7  
$\mu$m color image where the PAH emission (red)  traces both H$_{\alpha}$ and also the diffuse dust lanes.
A careful comparison of Figures 1 and 2 shows that in most cases the regions with strong 
7.7 $\mu$m emission also have strong H$_{\alpha}$ emission, a result that is established more quantitatively in \citet{hassani23}.

NGC~628 was observed in five bands by HST (WFC3/F275W, WFC3/336W, ACS/F435W, ACS/555W, and ACS/F814W) as part of the LEGUS survey \citep{calzetti15}, and was reduced using the PHANGS-HST pipeline (see \citet{lee22} for more details).  
Narrow-band observations of the H$_{\alpha}$ line taken with the ACS/F658N filter (Proposal 10402, PI: Chandar, Chandar et al. 2025 - submitted) also exist in the archive.
See \citet{lee22}\footnote{\url{https://archive.stsci.edu/hlsp/phangs-hst}}  for 
details about PHANGS-HST.
% PARA ON IMAGE ALIGNMENT AND FINAL PIXEL SCALE
The ACS data was drizzled onto a WFC3-like grid, using GAIA  \citep{gaia16} stars  for the alignment.
The pixel scale of 0.0396$\arcsec$ for WFC3 pixels corresponds to 1.890 pc~pix$^{-1}$ at the assumed distance of 9.84 Mpc for NGC~628. 

% JWST instruments, filters, and resolution
JWST observations of NGC~628 are available in the F200W, F300M, F335M, and F360M NIRCAM filters and in the F770W, F1000W, F1130W, and F2100W MIRI fiiters as part of a Cycle 1 JWST Treasury program (proposal 2107; PI: J. Lee; \citealt{Lee23}).
%%{\bf SUMMARIZE RESOLUTION FOR NIRCAM AND MIRI FILTERS.  Should we put basic information in a table ?}
See \citealt{Lee23} and \citealt{williams24} for additional details about the observations and basic reductions.

In this work, we start from the PHANGS-HST catalog of compact star clusters 
%% presented in 
\citep{whitmore21, thilker21a, Maschmann24}. ID numbers for the optically selected star clusters used in this paper are from \citet{Maschmann24}.  Clusters were selected to be extended sources which are broader than the PSF based on measurements of multiple concentration indices \citep{thilker21a}. 
% (the different in magnitude measured in two different size apertures).  
The extended sources in these initial catalogs were then classified by visual inspection by author B. Whitmore (human classification) and by machine learning algorithms into Class 1 (single-peaked, symmetric clusters), Class 2 (single-peaked, asymmetric clusters), Class 3 (multiple peak, compact associations) and Class 4 (contaminants) \citep{wei20,whitmore21,Hannon23}.  In this work we use the 489 human-classified Class $1+2$ catalog in NGC~628.  However, the number of clusters used in different parts of the analysis is smaller for a variety of reasons, including being faint (often in the UV) or missing due to a somewhat different field of view (primarily JWST filters).  Of the 489 clusters, 
%% 399 have U-B and V-I measurements which are required to put them on a color-color diagram used for several aspects of this paper, and 
only 291 have U-B, V-I, I-3.3 $\mu$m, and 3.0-7.7 $\mu$m measurements required to include the cluster in various parts of the analysis performed in this paper.

\subsection{Photometry}
%Photometry}
\label{sec:photometry}

\subsubsection{Small Versus Large Aperture Photometry }
%Photometry}
\label{sec:small_vs_large}

The difference in spatial resolution as a function of wavelength (e.g., effective radius of a PSF $\approx$ 0.08$^{\prime\prime}$
for the F555W filter  and $\approx$ 0.67$^{\prime\prime}$
for the F2100W filter) 
provides an important and difficult challenge for multi-wavelength  studies of star clusters. A common approach is to convolve the shorter wavelength images to have comparable resolution to the longer wavelength images, so that the apertures are effectively ``matched''. 
While this approach gives up some spatial resolution information, it gains a degree of uniformity in at least attempting to look at the same objects and field of view. 

In our case we are primarily interested in the star clusters, which are barely resolved and often very close together. Hence spatial resolution is our primary concern and we therefore use ``small aperture'' photometry as our primary method in most of this paper. In addition, it is not  always possible to ``look at the same objects", even when matching the field of view with convolved images and larger apertures. 
This is because the IR flux for the youngest regions (1 - 5 Myr) is generally dominated by emission from gas and dust (i.e., thermal dust continuum, PAH emission, warm ionized gas), while optical flux is dominated by stellar continuum emission. Hence, although the stellar 
light stays in one place and is observed in both the optical and IR, the gas and 
dust  associated with the cluster 
%% that generally dominates the flux in the IR, e
generally expands to sizes that are many times larger than even the large 0.67$^{\prime\prime}$ radius of the F2100W PSF in just a few Myr. 
Hence, it is not possible to actually match the features  associated with young clusters at all ages.

In Section \ref{sec:embedded} and Appendix B we return to this topic by making a comparison between small aperture photometry and photometry based on images convolved to the resolution of the 21 $\micron$ image based on the paper by  \citet{hassani23}. This 
provides  more quantitative information comparing the two approaches to performing photometry.
Luckily, the dynamic situation responsible for this problem also provides the solution for our age-dating goals, since it removes the gas and dust from the aperture in only a few million years, resulting in fluxes that vary by a factor of more than a hundred in the  F2100W filter in roughly five Myr.

\subsubsection{Photometric Parameters }
\label{inherent_problems}

Typically, an aperture radius of 0.1$^{\prime\prime}$ to  0.2$^{\prime\prime}$ (i.e., 3 to 5  ACS or WFC3 pixels using HST;  $\approx$ 5 - 10 pc  for galaxies at 10 Mpc) has been used to study star clusters in nearby galaxies \citep{whitmore10,chandar10a,bastian12b, adamo17}. This allows studies to  focus on a single star cluster in regions that are often very crowded while still excluding neighboring clusters and stars. In principle, we could use much larger radii that cover the entire HII complex,  often several hundred pc in size. However, this would result in the inclusion of several (or even dozens) of nearby star clusters covering a wide range of ages, hence defeating the basic aim of the project which is to study the ages and evolution of individual clusters. 

For the current paper, our primary method of performing photometry uses small apertures centered on  the clusters in the HST observations. More specifically,  a 4  WFC3-pixel (0.158$^{\prime\prime}$) radius with a sky annulus from  7 - 8 WFC3 pixels, and aperture corrections based on bright isolated star clusters (see \citealt{deger22} for details).  
We note that there are small (typically a few hundreds of a magnitude) differences between the HST photometry used in the current paper %(from  Rodriguez et al. 2025 - submitted) 
\citep[from][]{rodriguez24}
when compared to the original photometry in \citet{turner21}, due to the use of different software packages.

The spatial resolution in  NIRCAM bands is similar to HST, hence 
comparable 
4 pixel (0.124$^{\prime\prime}$) radii apertures were employed with a sky annulus from 7 - 8  NIRCAM pixels. Aperture corrections were derived based on bright isolated globular clusters in NGC 628.

Aperture photometry for the MIRI observations used the 50 \% encircled energy radii for stars. 
A sky annulus of 13 to 14 MIRI pixels 
(1.43$^{\prime\prime}$ and 1.54$^{\prime\prime}$) is used in all MIRI bands.   The total brightness is then obtained by doubling the measured flux, since the clusters are essentially point sources in the MIRI bands.  See \citet{rodriguez24} for details about the HST, NIRCAM, and MIRI photometry used in the current paper. This is referred to as ``small aperture photometry'' throughout this paper. 

It should be noted that while the aperture corrections are reasonable for the star clusters themselves, they are largely meaningless  for the ISM components associated with very young clusters, due to the rapid expansion of the superbubbles.  Fortunately,  as we will see in Section \ref{sec:templates}, there is a factor of $\sim100$ difference between the flux of a  1 and 5 Myr cluster in the MIRI bands, significantly larger than the uncertainties in the aperture corrections.

\begin{figure*}
\begin{center}
\includegraphics[width =7in, angle= 0]{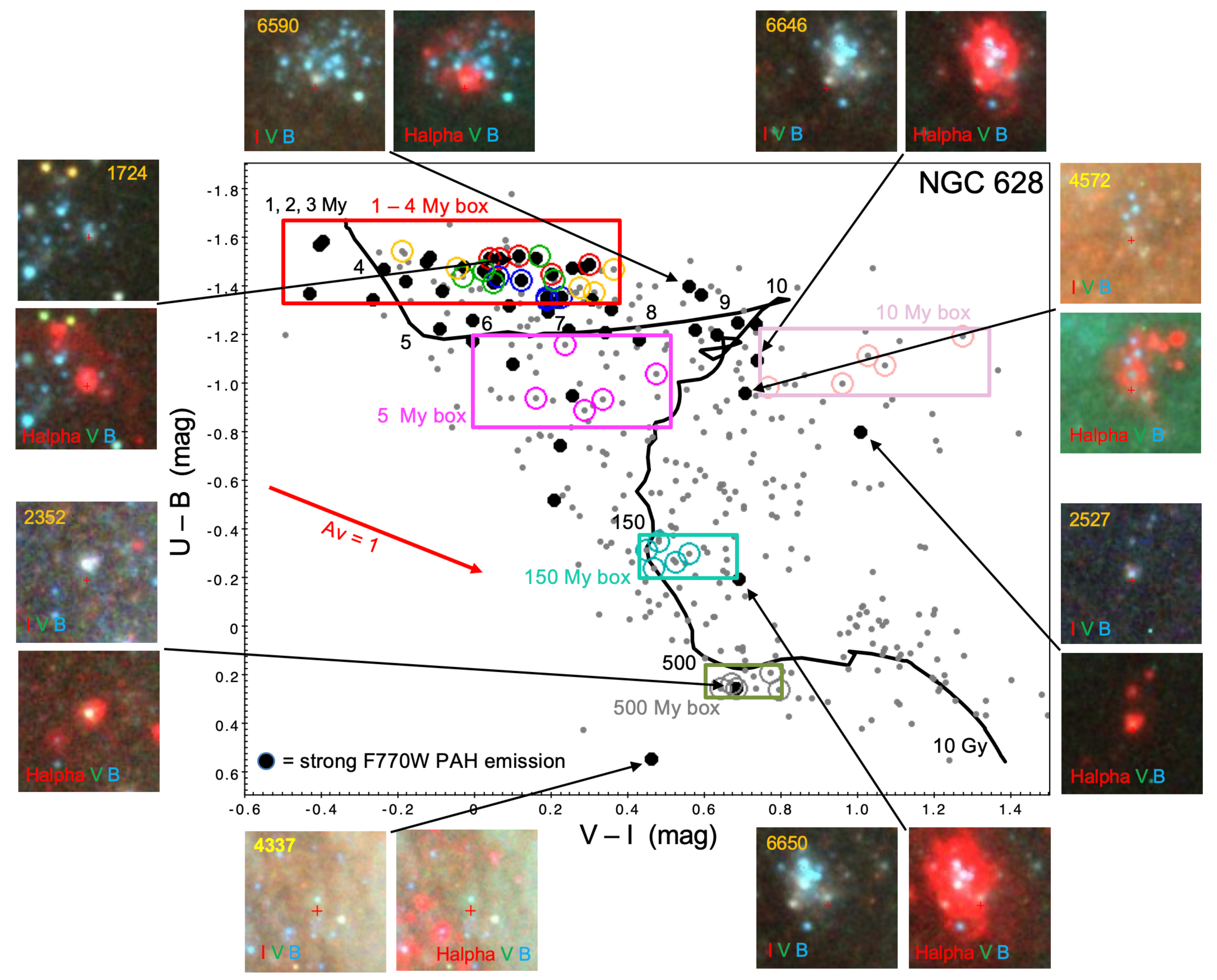}
% {pilot2_5_boxes_mar_13_2024.pdf}
\end{center}
\caption{ U-B vs V-I diagram with the five ``degeneracy-free" regions  shown, as discussed in the text. The clusters from the training set are identified using color-coded circles defined in Figure \ref{fig:image_and_ages_halpha}. Clusters with strong PAH emission (i.e., F770W / F300M $>$ 20) are shown in black. Most of these clusters with strong PAHS are above the 10 Myr position, as determined by the BC03 models shown by the black line, and are compatible with being from the 1 - 4 Myr box with varying degrees of reddening. Five of the six clusters which are potentially older than 10 Myr have strong H$_{\alpha}$ emission, as shown by the snapshots, and hence are young clusters (1 - 5 Myr) with  A$_V$ $\approx$ 1.2 - 2.5 mag. 
Hence, from this figure alone it is clear that nearly all strong PAH emitters have ages less than about 5 Myr.
}
\label{fig:cc_single_age}
\end{figure*}

Fluxes are measured/converted into Jansky for all filters.  When only optical colors are presented, they are converted to the VEGAMAG system\footnote{The VEGAMAG system has been used in most past studies which have focused on optical observations. The use of VEGAMAG for these particular figures is designed to  facilitate comparison to previous results such as \cite{chandar10a} and \cite{adamo17}.}
(for example Section~\ref{sec:non-degenerate} and Figure~\ref{fig:cc_single_age}), and when HST and JWST colors are plotted together, all filters are in the ABMAG system (for example in Section \ref{sec:color-color} and Figure \ref{fig:cc_ir}).

As discussed in Sections \ref{sec:small_vs_large}, \ref{sec:embedded}, and Appendix B,  we also perform convolved large aperture photometry and compare it with our results using small aperture photometry for some aspects of the analysis. Briefly, this uses ASTRODENDRO \citep{robitaille19} to identify sources in the F2100W image, and then measures fluxes at these locations on images of the  various filters which have been convolved to the same spatial resolution as the F2100W image (i.e., 0.67$^{\prime\prime}$). A background is subtracted using median values in an annulus which was 2 to 3 times larger than the F2100W resolution (i.e., 1.34 to 2.01$^{\prime\prime}$). See \citet{hassani23} for details. 

The photometry used in the current paper is the version available in summer 2024 from the various studies listed above. Later versions of the photometry may vary, generally slightly. For this reason the photometry is included in the tables of the current paper. 

\subsection{Cluster Age Estimates}
\label{sec:ages}

In this section we start with class 1$+$2 clusters identified in the human-classified PHANGS-HST catalog (see \citealt{Maschmann24} for details).
%for {\bf class 1 and 2 clusters}.
%% for which we can most reliably establish ages. 
We use two 
independent 
associated methods to estimates ages, both independent of the normal PHANGS SED age estimates reported in \citet{turner21}, Henny et al. 2025 (submitted), and  Thilker et al. 2025 (submitted). 
%% Thilker (2024 - in prep), and Henny (2024 - in prep)
\footnote{In principle, we could use ages determined by these other methods, but while most of these ages are robust, there are some remaining biases as well.}  

The first method relies on specific ranges of U-B and V-I colors which uniquely establish the ages of the clusters rather than being affected by the age/reddening/metallicity degeneracies.
This is possible for clusters with ages of $\approx $1 - 4, 5, 10, 150, and 500 Myr, which can then be used as a `skeleton' to build templates at all ages.

The second method uses the size of H$_{\alpha}$ superbubbles around very young clusters to approximately refine cluster ages of 1, 2, 3, and 4~Myr (see \citealt{whitmore11}, \citealt{hannon22}, \citealt{pedrini24}).
We select the five most massive clusters (generally $>$ 3000 Msolar) at each  age to minimize the impact of stochasticity on the integrated flux measurements (e.g., \citealt{fouesneau10,fouesneau12,krumholz15,wofford16,hannon19,whitmore20,turner21}).
The locations of these clusters are shown in Figures \ref{fig:image_and_ages_halpha} and \ref{fig:image_and_ages_770}.

\begin{table*}
% \caption{Census of Massive Embedded Star Clusters in NGC 1365 and Related Sources}
 \caption{Properties of the Cluster Training Sample Used to Make Empirical SED Templates }
 \label{tab:table_1}
 
 % \noindent\begin{tabular}{@{}lllllllllllll}
     \centering

\begin{tabular}{rrrrrrrrrrrrrrr}
  \hline
  & ID$^a$  & Adp. Age$^b$  & R$_{H\alpha}$$^c$  & RA & DEC  & SEDFIX age$^d$  & log M$^d$  & EBV$^{d,e}$  &  U-B$^e$  & V-I$^e$   & I-3.3$^f$  & 3.0-7.7$^f$     \\
 
%   & & & & & (J2000) & (J2000) &  &  &  & & &  &  \\
  & & (Myr) & (pix)  & (deg)  & (deg)  & (Myr)   & M$_{\odot}$  & (mag)  & (mag)   &  (mag)  & (mag) & (mag)  \\

\hline  

1 &  1724 & 1.0 & 0 & 24.160755 & 15.765572 & 2.0 & 3.49 & 0.26 & -1.52 & 0.12 & 1.07 & 4.29\\
2 &  1767 & 1.0 & 0 & 24.163475 & 15.765885 & 2.0 & 4.07 & 0.22 & -1.49 & 0.30 & 0.91 & 4.30\\
3 &  4505 & 1.0 & 0 & 24.160275 & 15.783140 & 2.0 & 3.62 & 0.10 & -1.45 & 0.20 & -0.10 & 4.45\\
4 &  5418 & 1.0 & 0 & 24.152300 & 15.789327 & 3.0 & 3.75 & 0.18 & -1.51 & 0.04 & 0.24 & 4.24\\
5 &  7280 & 1.0 & 0 & 24.176468 & 15.804831 & 3.0 & 3.53 & 0.14 & -1.51 & 0.07 & 0.42 & 4.69\\
6 &  3150 & 2.0 & 2 & 24.185388 & 15.775613 & 3.0 & 4.08 & 0.24 & -1.34 & 0.19 & -0.80 & 4.77\\
7 &  3538 & 2.0 & 2 & 24.185787 & 15.777906 & 3.0 & 3.83 & 0.18 & -1.44 & 0.06 & -0.02 & 3.93\\
8 &  4995 & 2.0 & 1 & 24.171566 & 15.786123 & 3.0 & 3.41 & 0.20 & -1.35 & 0.23 & -0.17 & 4.83\\
9 &  5015 & 2.0 & 2 & 24.161094 & 15.786205 & 3.0 & 3.51 & 0.20 & -1.35 & 0.19 & -0.01 & 3.70\\
10 &  6681 & 2.0 & 1 & 24.163612 & 15.799347 & 3.0 & 3.78 & 0.14 & -1.42 & 0.12 & -0.42 & 4.77\\
11 &  1905 & 3.0 & 3 & 24.174276 & 15.766726 & 3.0 & 3.39 & 0.10 & -1.41 & 0.05 & -0.99 & 4.48\\
12 &  1343 & 3.0 & 4 & 24.163696 & 15.763633 & 1.0 & 4.26 & 0.22 & -1.43 & 0.21 & -1.81 & -1.07\\
13 &  2972 & 3.0 & 4 & 24.187976 & 15.774374 & 3.0 & 4.22 & 0.10 & -1.44 & -0.03 & -1.71 & 2.31\\
14 &  4768 & 3.1 & 2 & 24.171274 & 15.784595 & 2.0 & 3.75 & 0.16 & -1.52 & 0.17 & -0.99 & 0.97\\
15 &  5434 & 3.0 & 3 & 24.152246 & 15.789426 & 3.0 & 3.69 & 0.18 & -1.46 & 0.02 & -1.78 & 4.60\\
16 &  1214 & 4.0 & 12 & 24.196543 & 15.762902 & 7.0$^g$ & 3.99$^g$  & 0.01$^g$ & -1.37 & 0.31 & 0.4 & -0.20\\
17 &  1248 & 4.0 & 5 & 24.195676 & 15.763044 & 3.0$^g$ & 3.60$^g$ & 0.03$^g$ & -1.54 & -0.19 & -3.14 & 1.24\\
18 &  5655 & 4.0 & 5 & 24.161463 & 15.790765 & 3.0 & 4.05 & 0.14 & -1.47 & -0.05 & -2.06 & 3.23\\
19 &  6276 & 4.0 & 17 & 24.163632 & 15.795263 & 1.0 & 4.68 & 0.24 & -1.39 & 0.27 & -0.09 & -0.85\\
20 &  7255 & 4.1 & 17 & 24.175683 & 15.804612 & 1.0 & 3.72 & 0.26 & -1.47 & 0.36 & -3.85 & 1.98\\
21 &  1667 & 5.0 & 5 & 24.164027 & 15.765316 & 1.0 & 3.84 & 0.34 & -1.16 & 0.23 & -2.32 & 2.55\\
22 &  2416 & 5.0 & 20 & 24.186201 & 15.770045 & 8.0 & 4.02 & 0.16 & -1.04 & 0.47 & -0.21 & -1.91\\
23 &  3433 & 5.0 & 6 & 24.185832 & 15.777333 & 5.0 & 3.61 & 0.28 & -0.94 & 0.16 & -1.42 & 2.31\\
24 &   5894 & 5.0 & -999 & 24.156443 & 15.792100 & 7.0 & 3.70 & 0.14 & -0.93 & 0.33 & -1.3 & -1.72\\
25 &   6895 & 5.0 & -999 & 24.171399 & 15.800950 & 4.0 & 3.74 & 0.46 & -0.89 & 0.29 & -3.06 & 0.53\\
26 &   1187 & 10.0 & -999 & 24.198491 & 15.762762 & 9.0$^g$ & 4.07$^g$ & 0.38$^g$ & -1.0 & 0.96 & -0.69 & -0.95\\
27 &  2688 & 10.0 & -999 & 24.169986 & 15.771925 & 10.0 & 4.07 & 0.12 & -1.11 & 1.03 & -0.43 & -1.76\\
28 &  4356 & 10.0 & -999 & 24.183245 & 15.782235 & 10.0 & 3.67 & 0.18 & -1.19 & 1.27 & 0.19 & -2.74\\
29 &  5016 & 10.0 & -999 & 24.154924 & 15.786206 & 10.0 & 3.90 & 0.14 & -1.07 & 1.07 & -0.43 & -3.76\\
30 &  7150 & 10.0 & 26 & 24.172343 & 15.803480 & 8.0 & 3.82 & 0.34 & -0.98 & 0.76 & -1.71 & 1.61\\
31 &  1743 & 150.0 & 15 & 24.161103 & 15.765697 & 124.0 & 4.04 & 0.10 & -0.30 & 0.56 & -3.42 & 0.90\\
32 &  2535 & 150.0 & 14 & 24.190482 & 15.770641 & 175.0 & 4.01 & 0.04 & -0.31 & 0.45 & -1.70 & 1.74\\
33 &  4901 & 150.0 & -999 & 24.184198 & 15.785612 & 175.0 & 4.03 & 0.06 & -0.35 & 0.48 & -1.52 & -0.29\\
34 &  5736 & 150.0 & -999 & 24.180184 & 15.791151 & 197.0 & 3.98 & 0.02 & -0.26 & 0.52 & -1.70 & -0.33\\
35 &  6400 & 150.0 & 6 & 24.161230 & 15.796543 & 156.0 & 4.33 & 0.00 & -0.24 & 0.47 & -1.33 & -2.05\\
36 &  1516 & 500.0 & -999 & 24.164762 & 15.764410 & 311.0 & 4.11 & 0.04 & 0.24 & 0.67 & 0.53 & -1.13\\
37 &  2273 & 500.0 & -999 & 24.165669 & 15.768983 & 391.0 & 4.57 & 0.06 & 0.26 & 0.64 & -1.38 & 0.94\\
38 &  2352 & 500.1 & 3 & 24.182627 & 15.769619 & 5.0 & 4.34 & 0.96 & 0.26 & 0.68 & -0.60 & 3.45\\
39 &  2984 & 500.0 & -999 & 24.162251 & 15.774448 & 874.0 & 4.48 & 0.08 & 0.26 & 0.80 & -1.29 & -2.02\\
40 &   6340 & 500.0 & -999 & 24.178579 & 15.796022 & 695.0 & 4.39 & 0.00 & 0.19 & 0.77 & -0.23 & -0.72\\

  \hline
  
 \hline
 
\end{tabular}

 \raggedright
 
$^a$ Star Cluster ID \# from \citet{Maschmann24}.\\
$^b$ Adopted age as discussed in Sections \ref{sec:ages}. The values with ``.1" attached  are the outliers discussed in Section \ref{sec:outliers}. 
%% These were not used in the determination of the empirical templates, and are not included in many of the figures.
\\
$^c$ Distance in HST pixels from star cluster to the nearest H$_{\alpha}$ feature., which may or may not be associated with the star cluster in the older objects.  
% Note that in many cases the bubble is incomplete and not centered on the cluster. 
Values of -999 are used when there is no H$_{\alpha}$ present in the snapshot.  \\
$^d$ - The SED-TreeFit (SEDFIX in heading) age, mass, and E(B-V) reddening estimates based on HST observations from Thilker et al.(2025 -submitted). \\
$^e$ - Values in VEGAmag. \\
% $^f$ -  Values used in calculation in Janskys.\\
$^f$ -  Values in ABmag.\\
$^g$ From \citet{turner21} since ages for these objects are missing in Thilker et al. 2025 (submitted).

\end{table*}

\subsubsection{Initial Age Estimates from Degeneracy-free Regions in the U-B vs. V-I Color-Color Diagram}
%\subsection{Initial Selection from non-degenerate-Regions in Color-Color Space}
\label{sec:non-degenerate}

Figure \ref{fig:cc_single_age} shows a U-B vs V-I color-color diagram for star clusters in NGC~628. This diagram is the work horse for several PHANGS-HST studies; the reader is referred to papers from the PHANGS collaboration \citep[e.g.,][]{whitmore21,deger21,turner21,Maschmann24}, as well as earlier papers such as \citet{whitmore02,chandar10a, adamo17}  for details.
While the degeneracy between age and reddening in broad-band colors can make it challenging to estimate the ages of clusters in many parts of this diagram (\citealt{anders04a,whitmore02,whitmore23b}, Thilker et al. 2025 - submitted), 
%% Thilker 2024 - in prep), 
there are regions in the color-color diagram where clusters can generally only have a single age or very small range of ages. We will refer to these as ``degeneracy-free regions", and the associated ages will be referred to as ``adopted" ages.  These regions are identified by the boxes in Figure \ref{fig:cc_single_age}. 

\begin{figure*}
\begin{center}
\includegraphics[width =5in , angle= 0]{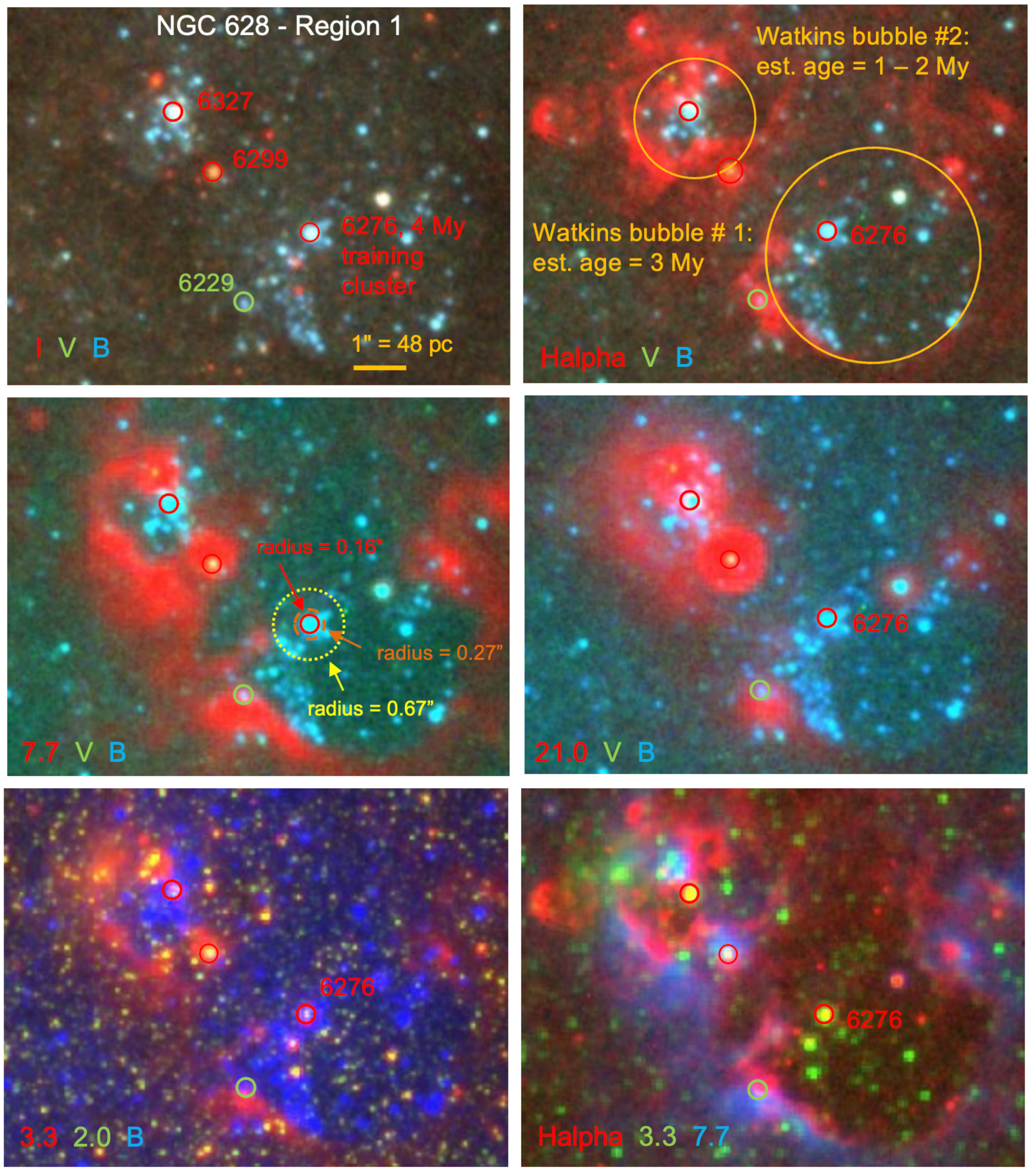}
\
\end{center}
\caption{Images using several color combinations are shown of Region 1 in NGC 628 (from Whitmore et al. 2025, in preparation) which contains two superbubbles from \citet{watkins23} (i.e., the yellow circles in the upper right panel) where age estimates have been made based on the observed expansion velocities in CO. 
% The larger superbubble has a radius of roughly 50 pc, which would result from an expansion velocity of 10 km/s for a period of 5 Myr (see text for discussion). 
There are four  clusters from the human-classified compact cluster catalog \citep{Maschmann24} (Class 1 in red, Class 2 in green), one of which (6276) is one of the 4 Myr training clusters used in the current paper.
The sizes of the apertures used in the optical (red), 7.7 (orange), and 21.0 (yellow) $\mu$m observations are shown in the middle left panel.  
}
\label{fig:region_1}
\end{figure*}

The box containing clusters with ages between 1 and 4 Myr  is shown in red 
% {\bf RC: change to blue?} 
and is the clearest example of a specific region in color-color space which is not very sensitive to the age-reddening degeneracy. The colors in this box are consistent with colors predicted by the \citet{bruzual03} (hereafter = BC03) solar-metallicity models for 1, 2, 3, and 4~Myr star clusters, with a small to moderate amount of reddening added 
(i.e., E(B-V) = 0.1 - 0.3 mag, hence  A$_V$ $\approx$ 0.3 - 0.9 mag). 
These clusters are blueward in U - B of the horizontal portion of the BC03 model line expected for 5 to 10 Myr clusters, when red supergiants (RSGs) start to appear (see Figure \ref{fig:cc_single_age}). Hence the clusters would require a negative (non-physical) value of $A_V$ to reach positions from the BC03 models with ages 6 to 9 Myr. 
A visual examination of all clusters in the 1 - 4~Myr box labeled in Figure \ref{fig:1_4_templates} shows that most 
have strong H$_{\alpha}$ emission, as expected for very young clusters. We take a closer look at the clusters in the 1 - 4~Myr box in Appendix A.

Table \ref{tab:table_1} shows that all the clusters in the 1 - 4~Myr box  with measured age estimates using the SED-TreeFit approach from Thilker et al. 2025 (submitted) have ages  less than 4 Myr.
This provides an independent sanity check that all of the clusters in the 1 - 4 Myr box are likely to be very young.

Similar degeneracy-free regions in the U-B vs V-I color-color diagram are identified for 4 - 7 Myr (maroon - 5 Myr box), 
9 - 11~Myr (gold - 10 Myr box), 100 - 250~Myr (cyan - 150 Myr box), and 300 - 800~Myr (grey - 500 Myr box). In all these cases, backtracking along the reddening vector (towards the upper-left) intersects the solar-metallicity BC03 model in a very small range of ages, hence minimising  degeneracies and providing more certain age estimates. This is not true for many other regions of the U-B vs V-I color-color diagram. For example, an old globular cluster can be assigned an age of 13 Gyr, 1 Gyr, 100 Myr, 6 Myr or 1 Myr, depending on how much reddening is assigned (see \citealt{whitmore23b}), Thilker et al. 2025 (submitted).

While different stellar evolutionary models, such as GALEV, Yggdrasil, STARBURST99, BC03 (\citealt{leitherer02, bruzual03, kotulla09, zackrisson11}) make somewhat different predictions for the evolution of U-B vs.\ V-I colors, we have found that the BC03 models provide a good overall match to the observed colors of $\sim100,000$ clusters from the PHANGS-HST survey (\citealt{turner21, Maschmann24}, Henny et al. 2025 (submitted), Thilker et al. 2025 (submitted).  Independent age estimates from spectroscopy and H$_{\alpha}$ bubble size also give similar results to the predictions from the BC03 models  \citep{whitmore11, chandar16, whitmore20}.

\subsubsection{Refined Age Estimates from H$_{\alpha}$ Suberbubble Sizes} 
\label{sec:bubbles}

% **RC: come back and clean up this section, since it describes 2 separate approaches to estimating ages from 'bubble' size. ***

\cite{whitmore11} demonstrated that the radius of H$_{\alpha}$ superbubbles around young clusters in M83 tends to increase with ages estimated from SED-fitting.
%correlate well with age estimates based on SED-fitting, bubble radius increases with cluster age. 
H$_{\alpha}$ emission is observed essentially on top of the broad-band stellar continuum emission for the youngest clusters; is in a small bubble for slightly older clusters; and has formed a fairly large bubble or shell (typically 20 - 80 pc) by an age of 3 - 5~Myr.
\cite{hollyhead15} performed a similar analysis in M83 and concluded that the clusters have removed their natal gas in $<$ 4 Myr, in good agreement with the \cite{whitmore11} timescale. Other studies with similar results include \citet{hannon22}, \citet{watkins23},   \citet{pedrini24}, and \citet{rodriguez24}.

Using these results, we  
divide clusters with integrated colors that fall in the 1-4~Myr box into 1, 2, 3, and 4~Myr adopted age bins based on their estimated H$_{\alpha}$ superbubble size.
Given 20 clusters, we 
%somewhat arbitrarily 
assign them into four groups of five each as described below. %% with ages 1, 2, 3, and 4 Myr. 
To facilitate this relative ordering by age we make simple visual estimates of the size of the superbubble, by measuring the distance in pixels from the cluster to the nearest H$_{\alpha}$ feature that appears to be part of the shell. The results are included in Table \ref{tab:table_1}, and are used to separate the sample into the appropriate age bins for the clusters in the 1-4~Myr box, also included in Table \ref{tab:table_1}. We note that most of the superbubbles are quite erratic in shape, making more complete measurement (e.g., with minimum and maximum shell radius as used in \citealt{hollyhead15} for typically better defined bubbles) unwarranted.

It is important to note that the adopted estimates in the 1 - 4 Myr age range should be considered as both tentative and relative (rather than absolute), since
there are many physical processes operating during the early life of a HII region that may modify the simple linear approximation used to assign age estimates for these very young clusters. %clusters into the adopted age bins}.  
Examples are inclusion of nebular emission in the models Thilker et al. 2025 (submitted), variations in escape fraction (e.g., \citealt{Barnes22,wei23}), coupling efficiency of stellar feedback (e.g., \citealt{weaver77,oey04,EGOROV_PHANGSJWST}),
modification of expansion velocities as a function of time \citep{watkins23}, age spread in the formation of low vs.\ high mass stars (e.g., \citealt{Brown22}), and different pressure of the ISM as a function of environment (e.g., as shown by a correlation of the sizes of HII regions with distance from the center of the galaxy; \citealt{pedrini24}). 
However, with a range of only 1 - 4 Myr for clusters in this part of the color-color diagram, essentially any reasonable scenario is going to give maximum errors of 1 or 2 Myr for a particular cluster. We take a closer look at the clusters in the 1 - 4 Myr box in Appendix A. One of the results  is the demonstration that there is a roughly even distribution of clusters from very compact to large bubbles, consistent with our assumption of an even distribution of ages from 1 to 4 Myr.

We note that nebular emission has not been included in the BC03 models shown in Figure \ref{fig:cc_single_age}. This will spread the predicted colors out slightly for 1, 2, 3 Myr ages, but the 1 - 4 Myr box is large enough that this will not affect the overall results very much (see Thilker et al. 2025 - submitted) for a discussion. 

\begin{figure*}
\begin{center}
\includegraphics[width =6.6in , angle= 0]{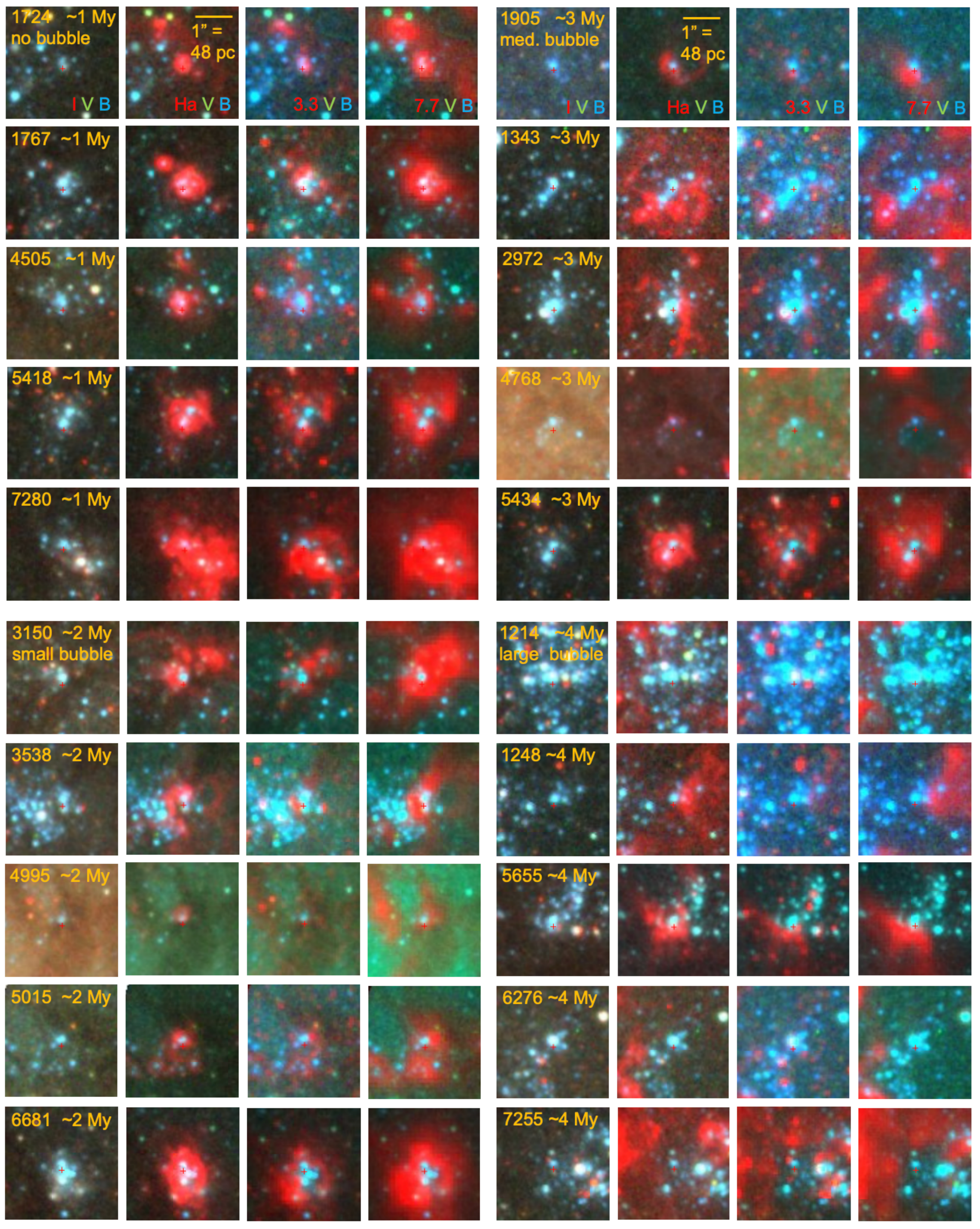}
\
\end{center}
\caption{Snapshots of  the 1 - 4 Myr training set clusters.  The cluster ID is shown in  yellow, along with the adopted age (see text). The color combinations  used for the snapshots, and a size scale, are shown in the top row. The program cluster is just above the small red cross in all cases. }
\label{fig:1_4_templates}
\end{figure*}

A sanity check is possible by examining the results from \cite{watkins23}, who identified 34 superbubbles in NGC 628, two of which are shown in Figure \ref{fig:region_1}. The bottom bubble (i.e., denoted Watkins - bubble \# 1) is \# 4 in the list of 12 ``perfect'' bubbles in NGC 628 in Table 2 of   \cite{watkins23}, with a CO  expansion velocity of 10 km/s and an estimated age of 3 Myr. The top bubble (i.e., denoted Watkins - bubble \# 2) is smaller and less resolved, and hence is not in the final catalog of 12 objects. However, Watkins' estimate for this object (private communications) is  1 - 2 Myr.
These age estimates are quite compatible with our use of bubbles to estimate cluster ages between 1 and 4~Myr.
%spread out the ages from 1 to 4 Myr. 
In particular, we note that the ``central'' cluster in the bottom superbubble is cluster 6276, which  is in our training sample with an adopted age of 4 Myr. Furthermore, the morphology of the top bubble is similar to the 2 and 3 Myr bubbles in Figure \ref{fig:1_4_templates}.

We also note that our estimated ages are consistent with those for young star clusters and HII regions in the Milky Way (e.g., \citealt{churchwell06}) and Magellanic Clouds. For example, \citet{demarchi11}, \citet{Crowther24} and \citet{fahrion24}  find ages in the range  2 to 4 Myr for various components of the central region of 30 Dor (i.e., R 136 and NGC 2070).

Figure \ref{fig:region_1} also demonstrates that the outflow that created the superbubbles has removed most of the gas and dust from within the effective photometric apertures used for our study (shown in the middle left panel as the red solid line for HST and NIRCAM data, the orange dashed line for the F770W data, and the yellow dotted line for the F2100W data) in just a few Myr. {\it It is primarily the rapid removal from the aperture of the gas and dust, and the corresponding nebular, PAH,  and IR dust emission, %% responsible for both the nebular and IR dust emission, 
that reduces the  near- and mid-infrared flux and produces the signal that allows us to 
% more accurately age-date
differentiate the youngest star clusters. }

\begin{figure*}
\begin{center}
\includegraphics[width =6.6in , angle= 0]{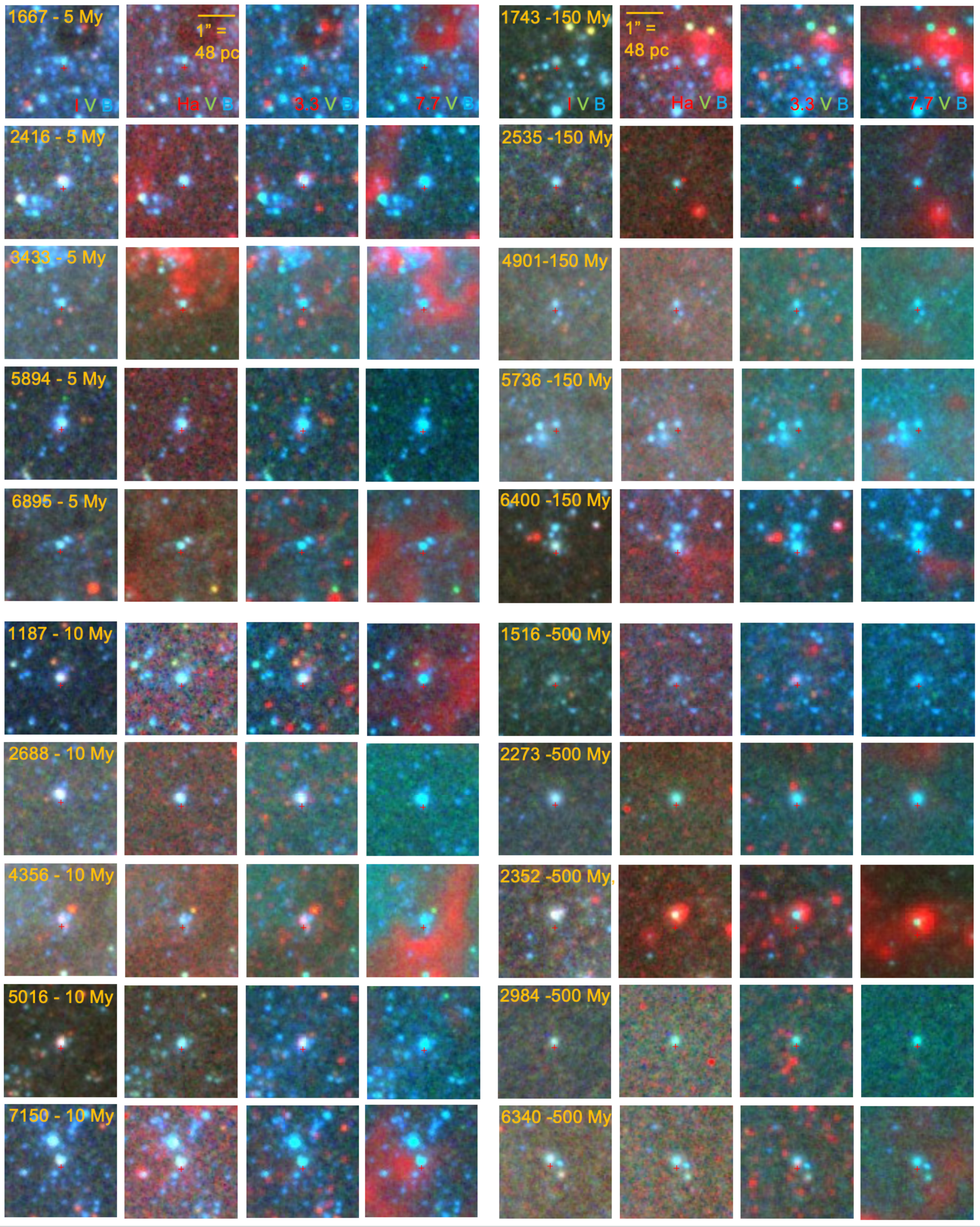}
\
\end{center}
\caption{Same as Figure \ref{fig:1_4_templates} for 5 - 500 Myr training set clusters.  }
\label{fig:5_500_templates}
\end{figure*}

\subsection{Selection of Clusters for SED Templates} 
\label{sec:training_set}

With our two associated age-dating methods in hand, we select the five most massive clusters (based on the Thilker et al. 2025 (submitted) estimates at each adopted age to create empirical SED templates.  Figures \ref{fig:1_4_templates} and \ref{fig:5_500_templates} show snapshots of all 40 clusters used for these templates. A careful examination of these images reveals several things. 

\begin{itemize}
    \item For the clusters with adopted ages  of 1~Myr clusters, H${\alpha}$ is right on top of the central cluster (always just above the small red cross). 
    \item The H$_{\alpha}$ superbubble 
    becomes less well defined from 3~Myr through 4~Myr, and H$_{\alpha}$ emission 
    is generally gone by 5~Myr. For older sources we do not generally see H$_{\alpha}$ clearly associated with the cluster in all but cluster 2352 (discussed further below as an outlier), just occasionally what appears to be unassociated diffuse  H$_{\alpha}$ emission based on the morphology (i.e., not ring-like  and not roughly centered  on the cluster). 
    \item H${\alpha}$ and 3.3 $\mu$m PAH emission have somewhat similar morphologies, although in many of the cases (e.g., sources 1767, 4505, 5418, 7280, 3538, 6681, 1905, 1343, 2972, 5434) the central cluster appears fainter in 3.3 $\mu$m, and the  superbubble appears slightly larger and/or the ring less-well defined than the H$_{\alpha}$ emission. This may be due to the preferential destruction of smaller PAH dust grains associated with 3.3 $\mu$m emission (e.g., \citealt{Madden06, povich07, maragkoudakis18, EGOROV_PHANGSJWST}). We return to this point in Section \ref{sec:temporal_profiles}.
    \item The 7.7 $\mu$m and 21 $\mu$m (not shown) maps have similar morphologies as the H$_{\alpha}$ maps,
    %% 3.3$\mu$m maps, 
    albeit with lower resolution. The resemblance between the 7.7 $\mu$m and 3.3 $\mu$m morphologies is not as good, again perhaps because of earlier destruction of the 3.3 $\mu$m  dust grains.

\end{itemize}
% The 5 Myr clusters show little associated H$_{\alpha}$.

We note that in one case (5418 and 5434) two clusters are part of a close pair, with a separation of about 10 pixels. As in previous studies (e.g., \citealt{whitmore21}) we use a separation of five WFC3 pixels as a criteria to reduce the redundancy and double counting. This is the only pair with a separation of less than 20 pixels in the sample, hence it has only a minimal effect on the results. 

The key takeaway from our investigation of the 40 clusters in NGC~628 that will be used to create SED templates is that by the time clusters are $\approx5$~Myr old, little-to-no associated H$_{\alpha}$, PAH or IR dust continuum emission remains. {The absence of strong gas and dust tracers in the 5 and 10~Myr samples, and the agreement of the observed IR SEDs with the TODDLERS predictions (discussed in Section \ref{sec:toddlers}), provide further support for the conclusion that there is little or no strong PAH or IR dust emission locally associated with clusters older than about 5 Myr. }

\bigskip

\subsection{Outlier Rejection}
\label{sec:outliers}

It is often the case that there are important lessons to be learned from outliers.  The small sample used in this pilot work allows us to study each cluster in detail, and identify objects to reject from the sample in order to reduce the scatter and improve the reliability.

A careful look at Figures \ref{fig:1_4_templates} and \ref{fig:5_500_templates} reveals three important issues that have been used to remove three of the clusters from our sample, as listed in Table \ref{tab:table_1}.
The first is related to the high background, and likely environmental dependencies of galaxies in the inner region of NGC 628. Cluster 4768 (3 Myr sample) is the primary example, showing a high background in all colors, and a potential environmental dependence caused by the inner HII regions being systematically smaller than HII regions in the outer part of the galaxy (e.g., see Figure  \ref{fig:image_and_ages_halpha}), probably because of the higher gas pressure in the inner region (e.g., \citealt{EGOROV_PHANGSJWST,pedrini24}). This cluster would have negative fluxes in the F1000W and F1130W filters if left in the sample.  A similar cluster found in the inner region is 4995 (2 Myr sample).
% as shown in Figures \ref{fig:image_and_ages_halpha} and \ref{fig:image_and_ages_770}. 
This second object in the inner galaxy has not been eliminated from the sample since the resulting photometry is not as discrepant when compared with the other clusters with age estimates of 2~Myr.

The second issue is relevant for cluster 7255
(4 Myr sample). This cluster has strong emission features in the outskirts which fall into the sky annulus (see Figure \ref{fig:1_4_templates}), resulting in the over-subtraction of all the emission features.  
This cluster would have  negative fluxes in the F335M and all four of the MIRI filters if left in the sample. 

The third issue is relevant for cluster 2352
(500 Myr sample). This is clearly a very young cluster as evidenced by the strong emission in H$_{\alpha}$, F335M and F770W. It is
probably in the ``wrong'' part of the color-color diagram due to stochasticity (i.e., the presence of one or more red super giants along with several blue stars). A careful look shows the presence of at least five point-like objects;  hence it is likely a  compact association rather than a cluster.
This cluster would have large positive flux measurements in all  MIRI bands, with values appropriate for a 2 or 3 Myr cluster, if left in the sample.

This is an important reminder that errors and noise from a variety of sources (e.g., classification errors, stochasticity, crowding, photometric errors, etc) are present, and may obscure the fundamental correlations if care (e.g.,  manual examination) is not taken into account when selecting a training cluster sample.

\begin{figure*}
\begin{center}
\includegraphics[width =6.6in , angle= 0]{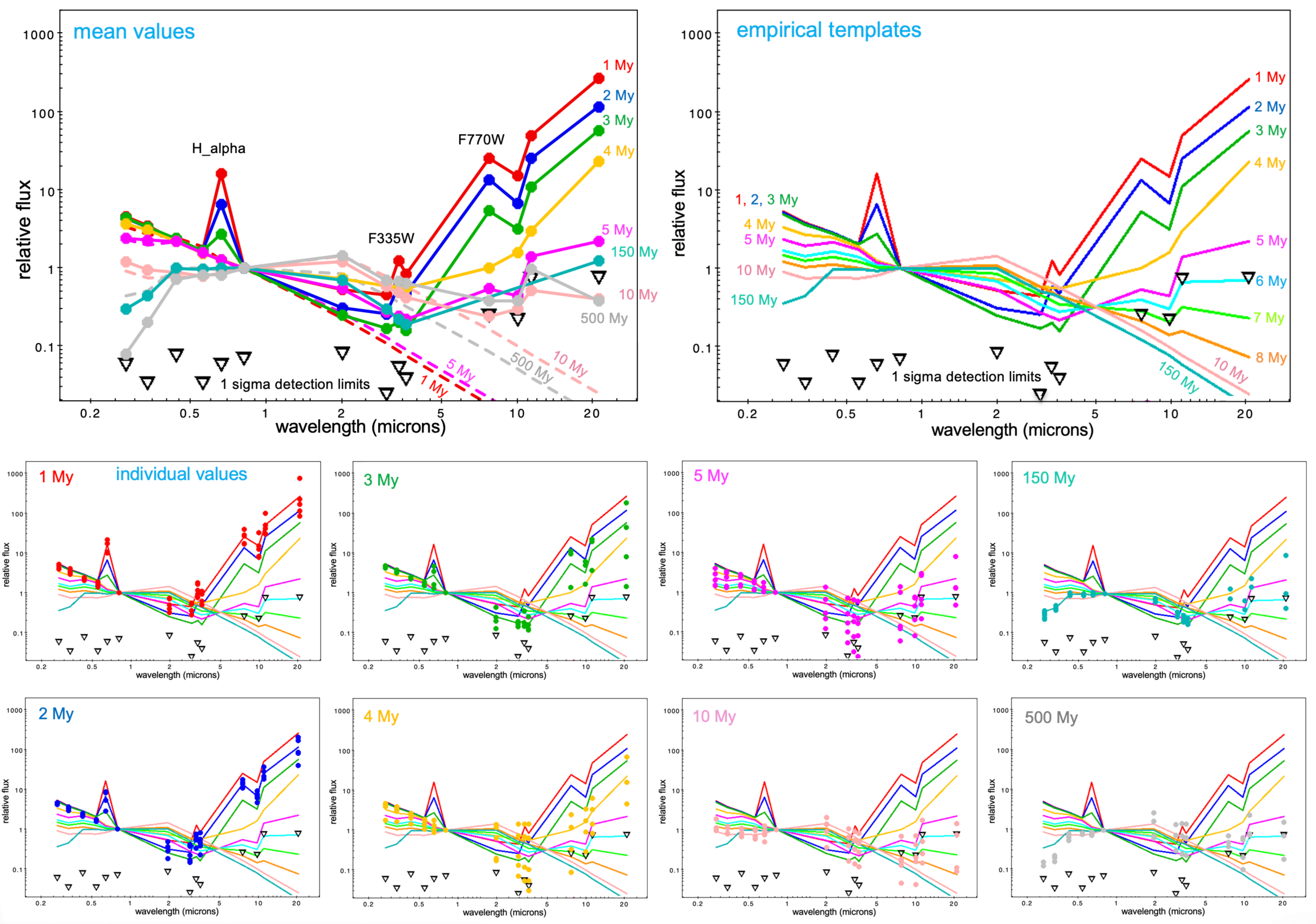}
\
\end{center}
\caption{Figure showing how the mean-value version of the  empirical templates (upper right panel) are derived from the observations of the training set (upper left panel). The dashed lines are from the BC03 stellar continuum models. The black triangles are the 1 sigma detection  limits. The individual  clusters are shown in the bottom panel for the five clusters in each sample, although in some cases there are fewer data points because a cluster is an outlier (see Section \ref{sec:outliers}), or is off the bottom of the diagram.}
\label{fig:mean_templates}
\end{figure*}

\section{Results: New Cluster SED Templates}
\label{sec:templates}

\subsection{Creation of Empirical Cluster SEDs}

We now use our measured photometry for the five most massive (generally brightest) clusters in each of the degeneracy-free boxes, coupled with information on the size and morphology of the HII region superbubbles, to create new empirical templates for 1, 2, 3, 4, 5, 10, 150, 500 Myr star clusters, and by extrapolation to  all cluster ages  since we will show that the observations for all clusters older than 10 Myr are compatible with BC03 stellar continuum models. Hence our templates for clusters older than  10 Myr are simply taken from the BC03 models.

We start by plotting the mean flux of the five clusters at each adopted age measured in each filter. These values are listed in Table \ref{tab:table_2}, along with the corresponding prediction from the BC03 solar metallicity, stellar continuum models in parenthesis for comparison. The three outliers discussed in the last section are not included. %% (but note a few caveats at the end of this section since 3 of the 40 clusters are later removed from the sample).  
The mean fluxes for each adopted age are plotted in the top-left panel of Figure~ \ref{fig:mean_templates}, where they have been normalized to 1 in the F814W (I) band. 
The fluxes are corrected for a small amount of reddening for each cluster (see Table \ref{tab:table_1}), based on the original HST fits from \cite{turner21}.
The lower panels show the measured fluxes for all five (four in the three cases where an outlier has been removed) clusters in each age bin to give a sense of the scatter. 

A great deal of insight can be
obtained by carefully examining Figure \ref{fig:mean_templates}. 
Here, we note some important features from the mean flux distributions that will be  important for building  our empirical templates.

In the optical bands using Hubble data (i.e., the six left-most data points from the F275W, F336W, F435W, F555W, F658N = H${\alpha}$, and F814W filters), the mean values agree with the BC03 stellar continuum models (i.e., the dashed lines - see also the numbers in parenthesis in Table \ref{tab:table_2}) remarkably well in almost all cases, providing strong support for both the age estimates of our training sets, and the use of the BC03 models. Only the 1, 2, and 3 Myr clusters show any  H${\alpha}$ emission that sticks out above the stellar continuum. 

Note that all of the data points in the optical part of the SED are well above the detection limits shown by the black triangles, hence these are high signal-to-noise observations. The detection limits are determined using the mean of the three lowest values of the standard deviations determined using the five clusters (or in some cases four) in  each of the six age-bins separately. Only the three lowest standard deviations are used to avoid bins with very large standard deviations (generally  the 4, 5, and 10 Myr bins).
%% due to stochasticity. 
% (see Table \ref{tab:table_1}).

In the near-IR bands using NIRCAM on JWST (i.e., F200W, F300M, F335M, F360M - the central four data points)
%% for the young clusters, 
the flux of the 1 and 2 Myr clusters are quite strong.
For the 3~Myr 
sample we find that the mean values drop significantly, and are essentially those expected from stellar continuum emission alone (i.e., the dashed lines).  This is likely because the small dust grains responsible for emission in this wavelength range have been removed by outflows, or destroyed by UV radiation, very quickly (e.g., \citealt{sandstrom12, EGOROV_PHANGSJWST, sutter24}). 
For ages of 5 Myr and older, red supergiants start appearing and the near-IR flux recovers and is much higher, reaching a peak at about 10 Myr where the flux in the NIRCAM filters is as strong as the brightest PAH emission in F335M for 1 Myr clusters! In particular, note the generally good agreement between the predicted BC03 stellar continuum and the data points for 10, 150, and 500 Myr clusters. 
{\it This is an important point; the only time that the NIRCAM flux is dominated by PAH or IR dust emission is for the 1 and 2 Myr clusters; in older clusters the near-IR flux is primarily from the stellar continuum.}  

The scatter between the points in the near-IR bands is quite small for ages 1, 2 and 3 Myr, but is much larger for ages 4 and 5 Myr. This probably reflects the irregular and patchy morphology of the older bubbles, as shown in Figures \ref{fig:1_4_templates} and \ref{fig:5_500_templates}. While many of the data points for the 4 and 5 Myr cluster are consistent with the stellar continuum values for ages of around 5 Myr  (as shown by the dashed lines in the upper left panel) other points are higher, and are more consistent with the predictions for ages of 7 to 10 Myr. It is not clear whether this is due to more uncertain age estimates in this age range based on the BC03 models (see Henny et al. 2025 - submitted), larger differences in morphology for the clusters with ages around 4 Myr, or larger photometric uncertainties since many of these clusters are near the 1 sigma detection limits. 

\begin{table*}
 \caption{Mean Relative Flux Values of the Eight Age Bins and Comparison to BC03 using Small Aperture Photometry$^a$}
 \label{tab:table_2}
 
 % \noindent\begin{tabular}{@{}lllllllllllll}
     \centering

\begin{tabular}{rrrrrrrrrrrrrrr}
  \hline
Wave$^b$  & 1 My  & 2 My & 3 My & 4 My  & 5 My & 10 My & 150 My & 500 My    \\

  0.275 & 4.56 (5.265) & 4.34 (5.187) & 4.40 (4.881) & 3.67 (3.385) & 2.40 (2.324) &  1.20 (0.915) & 0.30 (0.357) &  0.08 (0.095)  \\
  
0.336 & 3.49 (3.827) & 3.32 (3.786) & 3.34 (3.620) & 3.06 (2.687) & 2.29 (1.966) &  0.95 (0.746) & 0.44 (0.441) &  0.20 (0.189) \\

0.435 & 2.24 (2.819) & 2.35 (2.810) &  2.38 (2.765) & 2.20 (2.455) & 2.18 (2.178) &  0.88 (0.771) & 0.99 (0.983) &  0.71 (0.732)  \\

0.555 & 1.66 (2.056) & 1.65 (2.059) & 1.65 (2.041) & 1.54 (1.881) & 1.57 (1.728) & 0.79 (0.748) & 0.96 (0.983) &  0.82 (0.848)  \\

0.658 & \textcolor{red}{16.20}$^{c}$  (1.314)  & \textcolor{red}{6.61} (1.317) & \textcolor{red}{2.74} (1.312) &  1.29 (1.260)  & 1.26 (1.204) &  0.86 (0.807) & 1.01 (0.918) &  0.82 (0.859) \\

0.814 & 1.0 & 1.0 & 1.0 & 1.0 & 1.0 & 1.0 & 1.0 & 1.0  \\

2.0 & \textcolor{red}{0.53} (0.211) & \textcolor{red}{0.31} (0.207) & \textcolor{red}{0.25} (0.205) & \textcolor{red}{0.73} (0.228) & \textcolor{red}{0.54} (0.268) & 1.21 (1.430) & 0.70 (1.006)  & 1.46 (1.072) \\

3.0 & \textcolor{red}{0.45} (0.101) & \textcolor{red}{0.26} (0.099) & \textcolor{red}{0.17} (0.098) & \textcolor{red}{0.59} (0.109) & \textcolor{red}{0.27} (0.130) &  0.58 (0.781) & 0.30 (0.510) &  0.69 (0.566) \\

3.35 & \textcolor{red}{1.25} (0.080) & \textcolor{red}{0.59} (0.078) & \textcolor{red}{0.20} (0.078) & \textcolor{red}{0.57} (0.087) & \textcolor{red}{0.24} (0.104) &  0.48 (0.650) & 0.22 (0.473) &  0.67 (0.508) \\

3.6 & \textcolor{red}{0.84} (0.069) & \textcolor{red}{0.48} (0.068) & \textcolor{red}{0.16} (0.067) & \textcolor{red}{0.55} (0.075) & \textcolor{red}{0.22} (0.090) &  0.42 (0.575) & 0.19 (0.433) & 0.64  (0.475) \\

7.7 & \textcolor{red}{25.4} (0.016) & \textcolor{red}{13.60} (0.016) & \textcolor{red}{5.36} (0.016) & \textcolor{red}{1.01} (0.018) & \textcolor{red}{0.54} (0.022) & 0.24  (0.164) & -0.02 (0.123) & 0.38  (0.134) \\

10.0 & \textcolor{red}{15.0} (0.009) & \textcolor{red}{6.77} (0.009) & \textcolor{red}{3.15} (0.009) & \textcolor{red}{1.59} (0.010) & \textcolor{red}{0.44}  (0.013) &  0.30 (0.101) & -0.05 (0.078) &  0.38 (0.092) \\

11.3 & \textcolor{red}{49.4} (0.007) & \textcolor{red}{25.10} (0.007) & \textcolor{red}{11.00} (0.007) & \textcolor{red}{2.99} (0.008) & \textcolor{red}{1.41} (0.010) &  0.52 (0.079) & -0.64 (0.597) & 0.98  (0.067) \\

21.0 & \textcolor{red}{266.0} (0.002) & \textcolor{red}{116.5} (0.002) & \textcolor{red}{57.6} (0.002) & \textcolor{red}{23.0} (0.002) & \textcolor{red}{2.20} (0.003) &  0.40 (0.025) & 1.23 (0.017) &  0.38 (0.013) \\

  \hline
  
 \hline
 
\end{tabular}

 \raggedright

\bigskip

 NOTE: 
 
$^a$ The values without parenthesis are the mean values (using small aperture photometry) for the clusters in the training set for the  eight age bins used as a skeleton to derive the SED templates at all ages (see Figure \ref{fig:mean_templates}). The values in parenthesis are the corresponding values from the solar metallicity, stellar continuum BC03 models for comparison. The values are ratios using Janskys, normalized to the F814W filter. Note that H$_{\alpha}$ and PAH (i.e., 3.3, 7.7, 11.3 $\mu$m ) values in this table have not been continuum subtracted, but have been continuum subtracted in Figures \ref{fig:temporal_profile}. The mean values have been corrected for reddening using the Milky Way extinction coefficients and the E(B-V) values in Table \ref{tab:table_1}. \\
$^b$ Approximate wavelength in $\mu$m.\\
$^{c}$ The red numbers are the only ones that have been changed in the final empirical  templates in Table \ref{tab:table_3} (mean-value version), based on observed enhancements over the BC03 models in H$_{\alpha}$, PAH and IR dust continuum emission in our training set.

\end{table*}

\begin{table*}
 \caption{Empirical Templates$^a$ using Small Aperture Photometry}
 \label{tab:table_3}
 
 % \noindent\begin{tabular}{@{}lllllllllllll}
     \centering

\begin{tabular}{rrrrrrrrrrrrrrr}
  \hline
Wave$^b$  & 1 My & 2 My & 3 My & 4 My  & 5 My & 6 My  & 7 My &  8 My & 9 My  & 10 My & 100 My & 500 My & 1 Gy$^c$   \\

  0.275 & 5.265 & 5.187 & 4.881 & 3.385 & 2.324 & 1.688 & 1.464 & 1.214 & 0.941 & 0.915 & 0.441 &  0.095 & 0.034\\
  
0.336 & 3.827 & 3.786 & 3.620 & 2.687 & 1.966 & 1.444 & 1.252 & 1.027 & 0.772 & 0.746 & 0.511 &  0.189 & 0.113\\

0.435 & 2.819 & 2.810 & 2.765 & 2.455 & 2.178 & 1.645 & 1.417 & 1.134 & 0.791 & 0.771 & 1.010 &  0.732 & 0.438\\

0.555 & 2.056 & 2.059 & 2.041 & 1.881 & 1.728 & 1.399 & 1.233 & 1.013 & 0.761 & 0.748 & 0.998 &  0.848 & 0.661 \\

0.658 & \textcolor{red}{16.2} & \textcolor{red}{6.61} & \textcolor{red}{2.74} & 1.260 & 1.204 & 1.101 & 1.030 & 0.927 & 0.819 & 0.807 & 0.934 &  0.859 & 0.824\\

0.814 & 1.0 & 1.0 & 1.0 & 1.0 & 1.0 & 1.0 & 1.0 & 1.0 & 1.0 & 1.0 & 1.0 &  1.0 & 1.0\\

2.0 & \textcolor{red}{0.53} & \textcolor{red}{0.31} & \textcolor{red}{0.25} & \textcolor{red}{0.73}, \textcolor{blue}{0.23}$^d$ & \textcolor{red}{0.54}, \textcolor{blue}{0.27} & \textcolor{red}{0.68}, \textcolor{blue}{0.467} & \textcolor{red}{0.87}, \textcolor{blue}{0.691} & \textcolor{red}{1.10}, \textcolor{blue}{1.045} & 1.394 & 1.430 & 0.845  & 1.072 & 1.243\\

3.0 & \textcolor{red}{0.45} & \textcolor{red}{0.26} & \textcolor{red}{0.17} & \textcolor{red}{0.59}, \textcolor{blue}{0.11} & \textcolor{red}{0.27}, \textcolor{blue}{0.13} & \textcolor{red}{0.35}, \textcolor{blue}{0.240} & \textcolor{red}{0.45}, \textcolor{blue}{0.366} & \textcolor{red}{0.59}, \textcolor{blue}{0.564} & 0.759 & 0.781 & 0.446 &  0.566 & 0.654\\

3.35 & \textcolor{red}{1.25} & \textcolor{red}{0.59} & \textcolor{red}{0.20} & \textcolor{red}{0.57}, \textcolor{blue}{0.09} & \textcolor{red}{0.24}, \textcolor{blue}{0.10} & \textcolor{red}{0.30}, \textcolor{blue}{0.196} & \textcolor{red}{0.39}, \textcolor{blue}{0.301} & \textcolor{red}{0.50}, \textcolor{blue}{0.468} & 0.632 & 0.650 & 0.367 &  0.508 & 0.584\\

3.6 & \textcolor{red}{0.84} & \textcolor{red}{0.48} & \textcolor{red}{0.16} & \textcolor{red}{0.55}, \textcolor{blue}{0.07} & \textcolor{red}{0.22}, \textcolor{blue}{0.09} & \textcolor{red}{0.28}, \textcolor{blue}{0.171} & \textcolor{red}{0.35}, \textcolor{blue}{0.264} & \textcolor{red}{0.44}, \textcolor{blue}{0.413} & 0.559 & 0.575 & 0.322 &  0.475 & 0.551\\

7.7 & \textcolor{red}{25.4} & \textcolor{red}{13.60} & \textcolor{red}{5.36} & \textcolor{red}{1.01} & \textcolor{red}{0.54} & \textcolor{red}{0.40} & \textcolor{red}{0.29} & \textcolor{red}{0.21} & 0.157 & 0.164 & 0.085 &  0.134 & 0.155\\

10.0 & \textcolor{red}{15.0} & \textcolor{red}{6.77} & \textcolor{red}{3.15} & \textcolor{red}{1.59} & \textcolor{red}{0.44} & \textcolor{red}{0.30} & \textcolor{red}{0.21} & \textcolor{red}{0.14} & 0.096 & 0.101 & 0.051 &  0.092 & 0.108\\

11.3 & \textcolor{red}{49.4} & \textcolor{red}{25.10} & \textcolor{red}{11.0} & \textcolor{red}{2.99} & \textcolor{red}{1.41} & \textcolor{red}{0.68} & \textcolor{red}{0.32} & \textcolor{red}{0.16} & 0.075 & 0.079 & 0.040 &  0.067 & 0.077\\

21.0 & \textcolor{red}{266.0} & \textcolor{red}{116.5} & \textcolor{red}{57.6} & \textcolor{red}{23.0} & \textcolor{red}{2.20} & \textcolor{red}{0.71} & \textcolor{red}{0.23} & \textcolor{red}{0.07} & 0.024 & 0.025 & 0.012 &  0.013& 0.015\\

  \hline
  
 \hline
 
\end{tabular}

 \raggedright

\bigskip

 NOTES: These empirical templates (using small aperture photometry) are based on a combination of the ``skeleton" templates  based on the observations of the training set (Table \ref{tab:table_2}), and the solar metallicity, zero reddening, stellar continuum BC03 models (black and blue numbers), as discussed in Section \ref{sec:templates},  
 The red numbers are the only ones that have been changed, based on observed enhancements in H$_{\alpha}$, PAH and IR dust continuum emission in our training set. The full template for all individual ages (i.e., filling in the BC03 stellar continuum values for older ages) is available at https://archive.stsci.edu/hlsp/phangs/phangs-cat .

$^a$ The values in this table are ratios using Janskys, normalized to the F814W filter. Note that H$_{\alpha}$ and PAH (i.e., 3.3, 7.7, 11.3 $\mu$m ) values in this table have not been continuum subtracted, but have been continuum subtracted in Figure \ref{fig:temporal_profile}. The values have been corrected for reddening using the Milky Way extinction coefficients and the E(B-V) values in Table \ref{tab:table_1} for the training set. \\
$^b$ Approximate wavelength in $\mu$m.\\
$^c$ After 1 Gyr, lower metallicity BC03 models should generally be used. See \citet{whitmore23b} and  Thilker et al. 2025 (submitted), who use 1/50 solar metalicity values, appropriate for old globular clusters. \\
$^d$ When two values are listed, the red values are for version 1 (mean-value template) while blue values are for version 2 (lower-limit template). See Section \ref{sec:templates} for discussion.

\end{table*}

We have developed two versions of the empirical 4 - 8~Myr cluster SED models to reflect their larger scatter in the near-IR; one using the mean values for 4 - 8 Myr clusters (version 1) and one using the stellar continuum models in this age range (version 2). For simplicity, only the mean models are shown in the SED empirical templates shown in Figure \ref{fig:mean_templates}; both versions of the templates are shown in Section \ref{sec:color-color} where we discuss the IR color-color diagrams. Table \ref{tab:table_3} includes both versions of the empirical templates, with the mean value version shown in red and the stellar continuum version in red and blue. 

For the mid-IR bands using MIRI on JWST (i.e., the rightmost four points), 
we find a rapid increase in flux as a function of wavelength for the 1 and 2~Myr  clusters, with the 1~Myr flux measurements being slightly higher at all wavelengths.  The scatter in the fluxes between the five clusters at each age is quite small ($\approx$ 50 \%), so the shapes of these templates are quite secure, and are in fact, quite similar. 
The 5 Myr data points are much lower, and only slightly above the detection limit in the MIRI bands. For older ages (i.e., 10, 150, 500 Myr), the MIRI flux values are below the detection limit in most cases (i.e., several of the points are negative and do not show up in the plots showing the five clusters in each age range in the bottom of Figure \ref{fig:mean_templates}).

Based on the mean values shown in the upper left panel of Figure \ref{fig:mean_templates}, and the various points outlined below, we create template SEDs 
as shown in the  upper right panel of Figure \ref{fig:mean_templates}.  This series of different age templates provide a preliminary but important benchmark/comparison set which can be used to estimate cluster ages and improve model predictions using a combination of HST and JWST data.  Here are the primary steps used to build the templates.

\begin{itemize}

    \item{For the optical/HST bands we use the solar-metallicity BC03 stellar continuum models for all ages and filters, except for the F658N (H$_{\alpha}$) filter for ages 1, 2, and 3 Myr, where we use the mean observed values.}
  
    \item {For the NIR/NIRCAM bands we have two models, version 1 (mean-value template) using the mean values for ages 1 through 8 Myr, and version 2 (lower-limit template) using mean values  for ages 1 - 3 Myr but BC03 stellar continuum values for ages 4 to 8 Myr. } For all other ages we use the BC03 stellar continuum values.
    
    \item {For the MIR/MIRI bands we use the mean values for the 
    1, 2, 3, 4, and 5~Myr templates. 
    For the 6, 7, 8 Myr templates, since the 5 Myr data are barely above the brightness limits, we interpolate between the 5 Myr mean measured flux and the predicted BC03 stellar continuum for 9 Myr, in lieu of reliable measurements in this age range.  While this is somewhat ad hoc, it results in a relatively smooth evolution in the template SEDs from 1 to 9 Myr, and agrees with the stellar continuum position of the 10 Myr clusters in the IR-Optical color-color diagram   shown in Figure \ref{fig:cc_ir}, as we discuss in Section \ref{sec:color-color}}. 
  
    \end{itemize}

Table \ref{tab:table_3} provides the values for the two versions of the templates established in this paper. For both versions, most values are taken from the BC03 solar metallicity stellar continuum, $A_V = 0.0$ models. In version 1 (mean-value template = red values), the mean values from the training set are used in all cases.  For version 2, (lower-limit template = blue values) the mean (red) values are used for most of the values, but are superseeded by the blue values when appropriate (i.e., BC03 values are used for NIRCAM for ages 4 to 8 Myr).

Our full templates include 
%empirical 
predictions for all 14 HST$+$JWST filters presented here at all 109 ages (i.e., ranging from 1 Myr to 13.75 Gyr) included in the CIGALE implementation of the Bruzual \& Charlot (2003) models. This is done by filling in all values that are not red or blue in Table \ref{tab:table_3} with the stellar continuum values. 
The full template are available at \url{https://archive.stsci.edu/hlsp/phangs/phangs-cat}.

In principle we could attempt to construct SED templates using PHANGS clusters of all ages. This approach is complicated by the age/reddening/metallicity degeneracy which makes age estimates less certain outside of the degeneracy-free boxes we have used in this paper. Our approach to evaluate the SEDs using regions in the color-color diagram where we most trust the ages allows us to: 1) determine new empirical templates for the very youngest ages where the BC03 models do not 
%% necessarily 
fit the observed SEDs in the IR, and 2) establish that for 10, 150, 500 Myr the BC03 models agree with the observations relatively well. We therefore use the BC03 models for clusters of all ages greater than 10 Myr.

As appropriate for a pilot study, these SED templates should be considered preliminary. We expect it will be possible to build more accurate templates in the near future, both because of the coming availability of H$_{\alpha}$ observations for all 19 galaxies that have JWST observation (Chandar et al. 2025 -  submitted), and the inclusion of more clusters from more galaxies in our training set.

\begin{figure*}
\begin{center}
\includegraphics[width =6in , angle= 0]{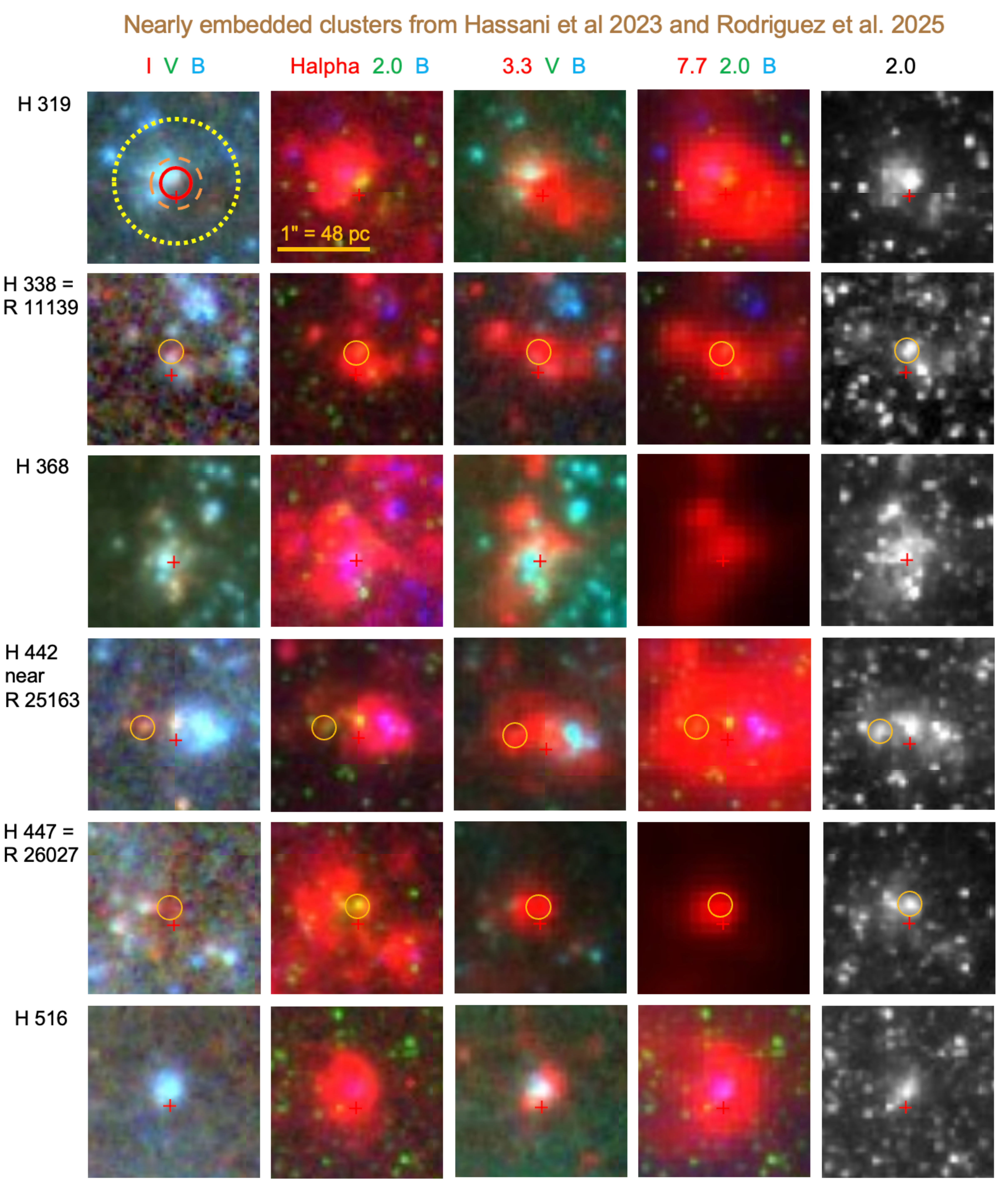}
\
\end{center}
\caption{{The first six of twelve nearly embedded cluster candidates from \citet{hassani23} (ID numbers starting with H and taken from the Hassani catalog). Four of the clusters are also either coincident with or near embedded cluster candidates from \citet{rodriguez24}, as shown by the small yellow circles (labeled with ID numbers starting with R and taken from the Rodriguez catalog).  See text for selection criteria and discussion.  The filters used to produce the images are indicated above  the top row. The photometric Field of View (hereafter = FOV) used for non-convolved (i.e., small aperture) photometry used for the training clusters and for \citet{rodriguez24} %% 2025  (submitted) 
photometry are shown in the first panel (i.e., red circles for optical and NIRCAM; orange dashed line for F770W; yellow dotted line for F2100W image). The large aperture convolved photometry used by \citet{hassani23} uses the yellow dotted line aperture for all wavelengths.  The scale is shown in the top row. 
The \citet{hassani23} program object is 5 pixels above the small red cross in all cases.}}
\label{fig:embed_mos_1}
\end{figure*}

\begin{figure*}
\begin{center}
\includegraphics[width =6in , angle= 0]{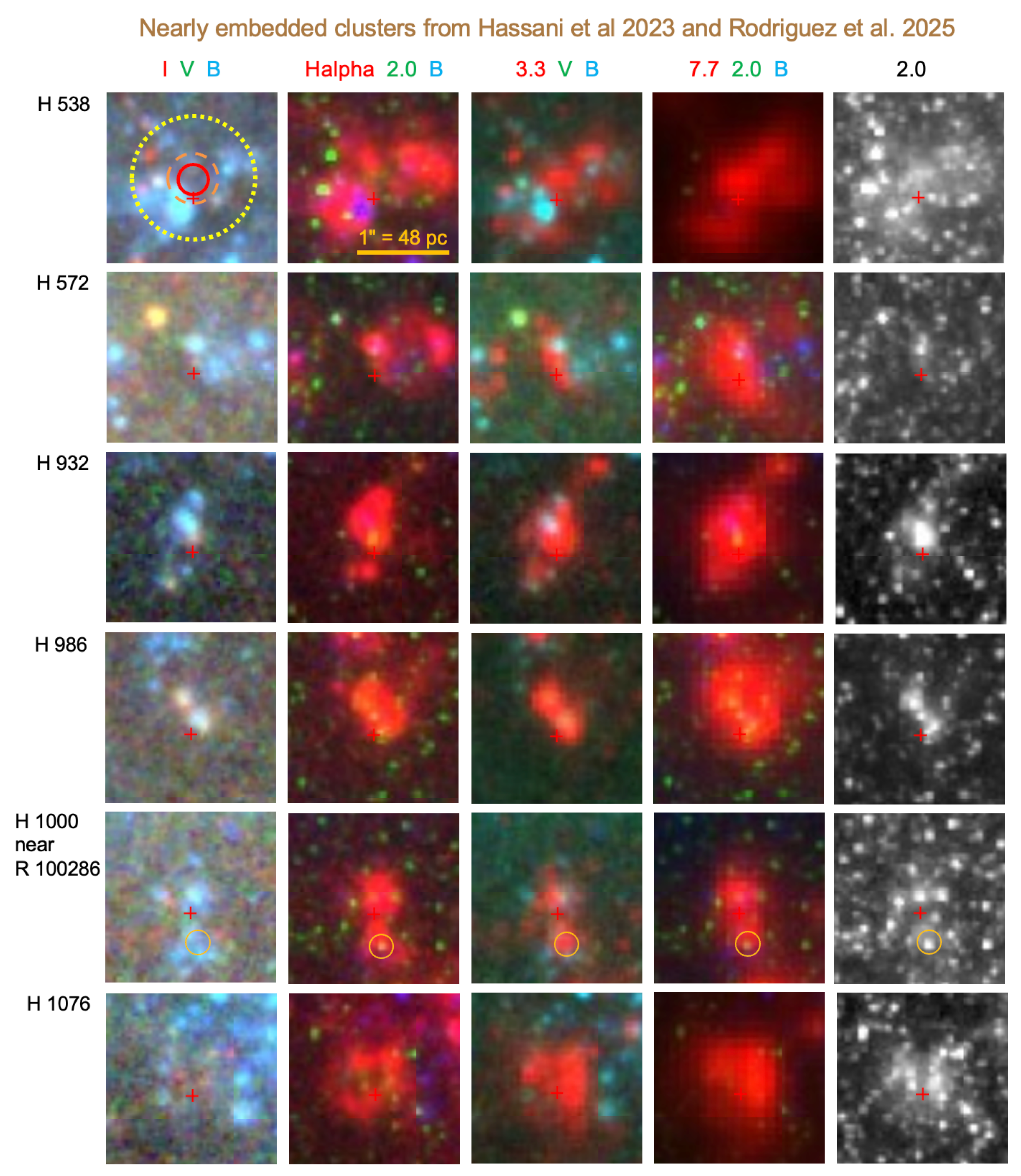}
\
\end{center}
\caption{Same as Figure \ref{fig:embed_mos_1} for the last six nearly embedded cluster candidates. }
\label{fig:embed_mos_2}
\end{figure*}

\subsection{Nearly Embedded Cluster Candidate SEDs}\label{sec:embedded}

\begin{figure}
\begin{center}
\includegraphics[width =3.3in , angle= 0]{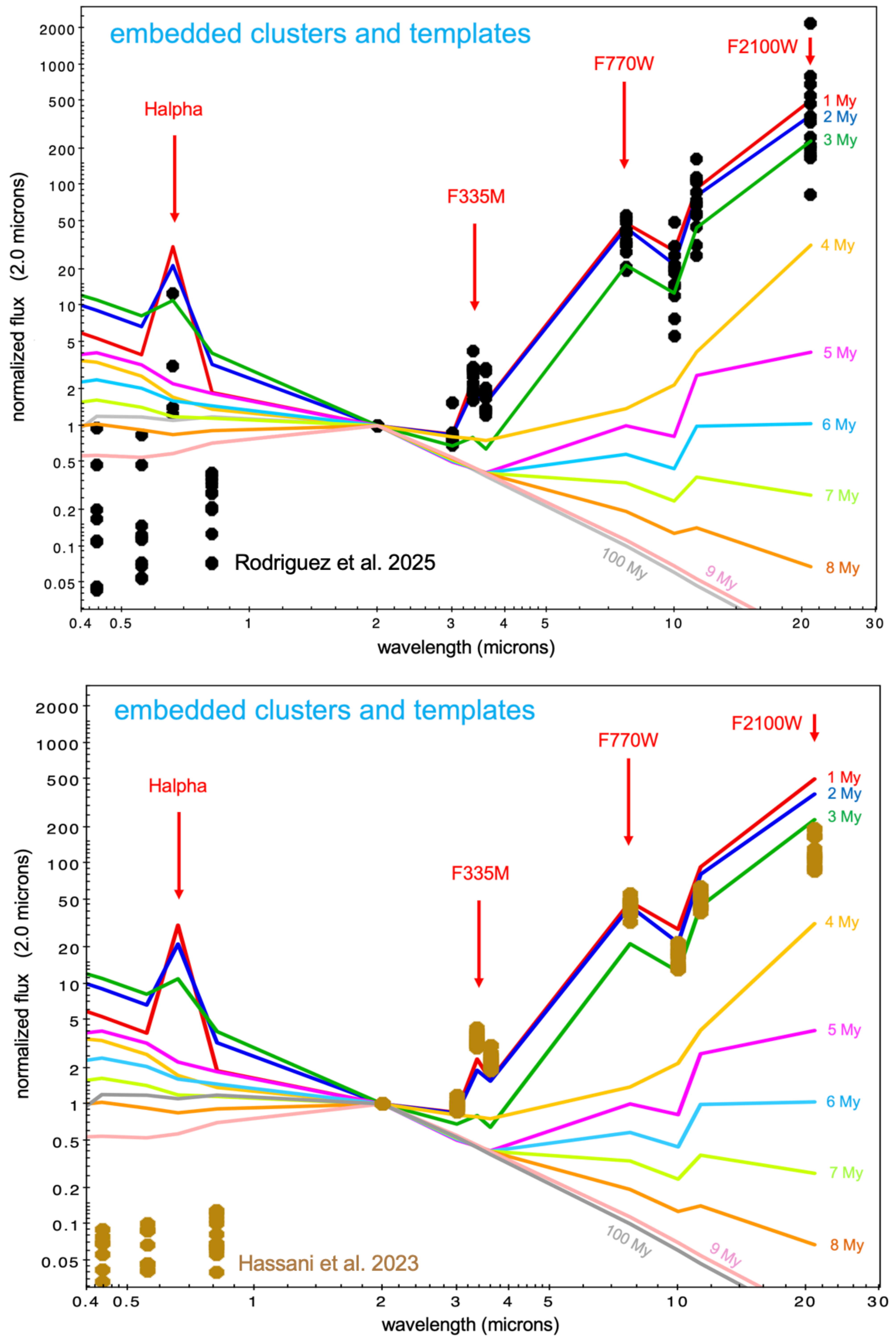}

\
\end{center}
\caption{SEDs for nearly embedded cluster candidates compared to empirical templates (mean-value version) based on optically identified clusters.  The top plot shows data from the \citet{rodriguez24} sample using small aperture photometry. The bottom plot shows data from the \citet{hassani23} sample using convolved large aperture photometry.
The same empirical templates as shown in Figure \ref{fig:mean_templates} are used, but are now normalized at 2.0 
$\micron$ rather than F814W, since several of the nearly embedded clusters are very faint 
%% or missing 
in the F814W filter.  We find that the nearly embedded clusters have SEDs similar to the 1 to 3 Myr clusters in the optical training sample, and the convolved photometry results in slightly shallower SED slopes in the mid-IR.
}
\label{fig:embedded}
\end{figure}

\begin{table*}
% \caption{Census of Massive Embedded Star Clusters in NGC 1365 and Related Sources}
 \caption{Normalized Flux Values for Nearly Embedded Cluster Candidates from \citet{hassani23}  Sample$^a$ Using Convolved Large Aperture Photometry}
 \label{tab:table_4}
 
 % \noindent\begin{tabular}{@{}lllllllllllll}
     \centering

\begin{tabular}{rllrrrrrrrrrrrr}
  \hline
 ID  & RA & DEC & 0.435 & 0.555  & 0.814 & 2.0 &  3.0  & 3.3 & 3.6 & 7.7 & 10.0 & 11.3 & 21.0 & 2.0-flux$^b$  \\

\hline

H 319 & 24.193756 & 15.758428 & 0.03 & 0.05 & 0.07 & 1.0 & 1.18 & 3.78 & 3.07 & 42.23 & 18.46 & 51.06 & 164.83 & 0.0237\\
H 338 &   24.177973 & 15.760403 & 0.01 & 0.01 & 0.04 & 1.0 & 0.99 & 3.55 & 2.25 & 46.48 & 16.88 & 53.06 & 112.53 & 0.0067\\
H 368 & 24.196670 & 15.763851 & 0.07 & 0.09 & 0.13 & 1.0 & 0.9 & 3.08 & 2.0 & 37.64 & 13.59 & 43.12 & 85.43 & 0.0256\\
H 442 &   24.196642 & 15.769243 & 0.06 & 0.07 & 0.12 & 1.0 & 0.91 & 3.25 & 2.21 & 45.12 & 21.46 & 57.13 & 191.79 & 0.0209\\
H 447 &  24.196194 & 15.769644 & 0.0 & 0.01 & 0.03 & 1.0 & 0.91 & 3.07 & 2.02 & 42.03 & 16.6 & 49.59 & 171.53 & 0.0193\\
%% 472 &  24.196129 & 15.770599 & 0.02 & 0.04 & 0.08 & 1.0 & 1.04 & 4.29 & 2.68 & 56.07 & 20.67 & 62.93 & 117.03 & 7.298302322971977\\
H 516 &  24.182970 & 15.773744 & 0.09 & 0.1 & 0.13 & 1.0 & 1.06 & 4.0 & 2.53 & 49.15 & 22.16 & 64.31 & 192.34 & 0.0091\\
H 538 &  24.185340 & 15.775685 & 0.07 & 0.09 & 0.1 & 1.0 & 1.0 & 3.45 & 2.28 & 48.04 & 14.85 & 51.05 & 119.11 & 0.0165\\
H 572 & 24.164656 & 15.777629 & 0.08 & 0.09 & 0.11 & 1.0 & 0.85 & 3.03 & 1.97 & 36.92 & 13.09 & 42.62 & 96.36 & 0.0069\\
H 932 &  24.149864 & 15.796673 & 0.03 & 0.04 & 0.06 & 1.0 & 0.99 & 4.0 & 2.5 & 49.95 & 19.38 & 57.56 & 131.78 & 0.0116\\
H 986 &  24.179390 & 15.800398 & 0.02 & 0.03 & 0.06 & 1.0 & 0.87 & 2.89 & 1.91 & 32.51 & 13.5 & 38.51 & 107.91 & 0.0159\\
%% 992 &  24.15329 & 15.800462 & 0.01 & 0.01 & 0.03 & 1.0 & 0.99 & 3.46 & 2.45 & 48.54 & 16.96 & 52.58 & 104.3 & 5.790922542241725\\
H 1000 &  24.179485 & 15.800799 & 0.04 & 0.05 & 0.06 & 1.0 & 1.03 & 3.38 & 2.1 & 39.51 & 14.22 & 44.93 & 92.26 & 0.0067\\
H 1067 &   24.179934 & 15.805174 & 0.01 & 0.02 & 0.06 & 1.0 & 0.93 & 3.2 & 2.33 & 40.81 & 13.1 & 42.44 & 90.14 & 0.0102\\

  \hline
  
 \hline
 
\end{tabular}

 \raggedright

\bigskip

$^a$ Selected nearly embedded cluster candidates from \citet{hassani23}, as shown in Figures \ref{fig:embed_mos_1} and \ref{fig:embed_mos_2}. See Section \ref{sec:embedded} for selection criteria. The flux values in this table are ratios using Janskys, normalized to the F200W filter. The values have not been corrected for extinction since we  do not have E(B-V) values from a SED fit using the HST data in most cases. \\
$^b$ The 2.0-flux in mJanskys can be used to  convert the relative flux values in this table to absolute values.

\end{table*}

\begin{table*}
 \caption{Normalized Flux Values of Nearly Embedded Cluster Candidates from \citet{rodriguez24} Sample$^a$ Using Small Aperture Photometry}
 \label{tab:table_5}
 
 % \noindent\begin{tabular}{@{}lllllllllllll}
     \centering

\begin{tabular}{rllrrrrrrrrrrrrrrr}
  \hline
 ID  & RA & DEC & 0.435 & 0.555  & 0.658 & 0.814 & 2.0 &  3.0  & 3.3 & 3.6 & 7.7 & 10.0 & 11.3 & 21.0 & 2.0-flux$^b$ \\

\hline  
 
%% H 1067 &   24.179934 & 15.805174 & 0.01 & 0.02 & 0.06 & 1.0 & 0.93 & 3.2 & 2.33 & 40.81 & 13.1 & 42.44 & 90.14 & 10.2E-12\\

 R 11139 & 24.177998 & 15.760442 & 0.11 & 0.12 & 0.0 & 0.4 & 1.0 & 0.79 & 2.94 & 1.84 & 31.81 & 14.95 & 55.42 & 212.06 & 0.00364 \\ %% & 2.15 & 4.00\\
 R 15153 & 24.196577 & 15.763497 & 0.47 & 0.47 & 12.52 & 0.27 & 1.0 & 1.56 & 3.03 & 2.98 & 19.19 & 5.57 & 25.81 & 165.2 & 0.00185 \\ %% & 2.61 & 2.72\\
 R 25163 & 24.196716 & 15.76925 & 0.02 & 0.07 & 3.14 & 0.37 & 1.0 & 0.69 & 1.91 & 1.34 & 38.92 & 31.03 & 108.86 & 790.02 & 0.00459 \\ %% & 1.78 & 4.38\\
 R 26027 & 24.19616 & 15.769671 & -0.01 & -0.01 & 1.25 & 0.07 & 1.0 & 0.82 & 2.87 & 1.92 & 40.87 & 20.08 & 69.04 & 371.0 & 0.00893 \\ %% & 3.99 & 4.24\\
 R 37861 & 24.16679 & 15.775292 & 0.2 & 0.0 & 0.0 & 0.21 & 1.0 & 0.87 & 2.26 & 1.81 & 31.71 & 14.18 & 57.59 & 197.7 & 0.00484 \\ %% & 2.58 & 3.90\\
 R 41648 & 24.170889 & 15.776963 & 0.04 & 0.05 & 0.0 & 0.13 & 1.0 & 0.71 & 2.04 & 1.4 & 27.78 & 12.06 & 44.9 & 176.78 & 0.00521 \\ %% & 3.01 & 3.977\\
 R 43865 & 24.185879 & 15.777877 & -0.34 & 0.05 & 0.0 & 0.32 & 1.0 & 1.54 & 4.18 & 2.79 & 56.2 & 26.06 & 113.77 & 680.95 & 0.00229 \\ %% & 2.80 & 3.90\\
 R 60799 & 24.170343 & 15.783778 & 0.17 & 0.11 & 0.0 & 0.28 & 1.0 & 0.79 & 3.03 & 2.04 & 42.21 & 18.47 & 74.15 & 331.07 & 0.00672 \\ %% & 2.57 & 4.31\\
 R 62243 & 24.170837 & 15.784217 & 0.95 & 0.83 & 0.0 & 0.4 & 1.0 & 0.71 & 1.63 & 1.23 & 20.5 & 7.75 & 31.43 & 82.93 & 0.00425 \\ %% & 1.52 & 3.64\\
 R 100286 & 24.184801 & 15.800811 & 0.01 & 0.01 & 0.0 & 0.2 & 1.0 & 0.77 & 2.57 & 1.69 & 34.11 & 21.58 & 67.5 & 247.47 & 0.00584 \\ %% & 2.77 & 4.11\\
 R 119112 & 24.196215 & 15.769671 & 0.04 & 0.07 & 1.42 & 0.21 & 1.0 & 0.86 & 2.82 & 1.9 & 51.95 & 30.66 & 104.42 & 543.62 & 0.00616 \\ %% & 2.83 & 4.45\\
%% R 119442 & 24.170282 & 15.783815 & 0.05 & 0.12 & 0.0 & 0.07 & 1.0 & 0.75 & 1.86 & 1.31 & 41.56 & 19.92 & 86.35 & 468.59 & 2.738390187254395E-4 & 3.5416359364071752 & 4.35\\
 R 119801 & 24.17634 & 15.804625 & 0.11 & 0.15 & 0.0 & 0.35 & 1.0 & 0.72 & 1.86 & 1.37 & 49.06 & 48.9 & 163.92 & 2180.65 & 0.00393 \\ %% & 1.81 & 4.57\\

  \hline
  
 \hline
 
\end{tabular}

 \raggedright

\bigskip

$^a$ Selected nearly embedded cluster candidates from \citet{rodriguez24}. See Section \ref{sec:embedded} for selection criteria. The flux values in this table are ratios using Janskys, normalized to the F200W filter. The values have not been corrected for extinction since we  do not have E(B-V) values from a SED fit using the HST data in most cases. \\
$^b$ The 2.0-flux in mJanskys can be used to  convert the relative flux values in this table to absolute values.

\end{table*}

With the optically-selected cluster templates in hand, we now check if nearly embedded clusters have similar SEDs in the infrared so they can be age-dated using the same empirical templates.

Samples of nearly embedded cluster candidates were obtained from two studies; \citet{hassani23} (using photometry convolved to match  the  0.67$^{\prime\prime}$ effective radius of the F2100W filter), and \citet{rodriguez24}, using the small aperture photometry described in Section \ref{sec:training_set} and used for our training set, empirical templates, and throughout the  rest of the paper. 

For the purposes of the current paper, the selection criteria designed to identify roughly a dozen of the best nearly embedded cluster candidates from the \citet{hassani23} sample was to have: 1) 2 $<$ F335M - F300M $<$ 3 (i.e., very strong F335W PAH emission) where the normalized flux values from Table \ref{tab:table_4} are used for the evaluation, 2) F200W flux  $>$ 0.006 mJy (i.e., strong continuum sources), and  3) normalized F814W between 0.0 and 0.13 (i.e., ``nearly embedded" in the optical). 

A similar set of selection criteria for nearly embedded cluster candidates from the \citet{rodriguez24} sample was used to select the 12 of 53 objects from the  5 sigma sample of 3.3 $\mu$m-emitters that have the following attributes: 1) F335M - F300M $>$ 0.9 (where the normalized flux values from Table \ref{tab:table_5} are used for the evaluation) , 2) F200W flux  $>$ 0.0015 mJy, and 3) normalized F814W between 0.0 and 0.4. See \citet{rodriguez24}
for details on how the original sample of  5 sigma, 3.3 $\mu$m emitters were selected.  We note that the  original selection of objects identified in the full Hassani catalog used the F2100W filter. The secondary selection using the F335M - F300M criteria was designed to better match the criteria used in the \citet{rodriguez24} catalog so that a fairer comparison could be made.

Figures \ref{fig:embed_mos_1}
and \ref{fig:embed_mos_2} show snapshots of the 12 \citet{hassani23} nearly embedded cluster candidates using several combinations of the filters to make the image, starting with the optical IVB image on the left. Four of these clusters are the same as, or nearby the \citet{rodriguez24} sample of nearly embedded cluster candidates, as shown by the yellow circles in the figure. The other \citet{rodriguez24}
clusters are not shown, but their photometric characteristics are listed in Table \ref{tab:table_5}.

A careful look at Figures \ref{fig:embed_mos_1} and \ref{fig:embed_mos_2} show that in  most cases (10 of 12)  the nearly embedded cluster can be seen as a very faint reddened object(s) in the IVB image. However, H 516 and H 1000 appear to be associated primarily with bluish optical sources, with no faint reddened objects associated with them.  

Another important result is that in nearly all cases (11 of  12), there is a relatively bright blue cluster within a few tenths of an arcsec. The only exception is  H~1076, and even then there are two bright clusters roughly 0.7$^{\prime\prime}$ away.  This suggests that the formation of most nearly embedded clusters was triggered by a young nearby (10 - 30 pc) star cluster. In four of the cases the nearby blue cluster is in the HST human-based cluster catalog \citep{Maschmann24}  as a class 1 or 2   object with a young age (i.e., H 368 - 4 Myr, H 516 - 3 Myr, H 538 - 4 Myr, and H 572 3 Myr), as determined by Thilker et al. 2025 (submitted).

In several cases a dark brick-like dust feature appears to be associated with the nearly embedded cluster candidate, as generally seen best in the IVB image. Perhaps the best examples of this morphology are H 319, H 368, H 442, H 447, H 538, H 572, H 932, H 1076 (especially the H$_{\alpha}$ image for this object).

Other things to notice are: 1) the objects are quite prominent in all three of the emission bands (somewhat less so in F335M) and generally have similar morphology that are most like the  1 and 2 Myr training objects in Figure \ref{fig:1_4_templates}, 2)  in the F200W filter the clusters are generally seen as fairly prominent, fuzzy objects, often with multiple point-like sources associated with them, 3) the contrast with the very crowded background around the central F200W source is quite low in many cases, implying that the uncertainty in the photometry may be larger than in most other filters (examples are H 338, H 572, and H 1000).

We show the SED flux measurements for both the \citet{hassani23} and \citet{rodriguez24} samples in Figure \ref{fig:embedded}, which is similar to Figure~\ref{fig:mean_templates} but now normalized at F200W since some of the nearly embedded clusters are not well detected in the F814W filter. These clusters have optical broadband  and H$_{\alpha}$ (only available for the Rodriguez sample) flux measurements   that are lower than the optically-based templates in Figure \ref{fig:embedded}, as expected for clusters with high extinction (which we cannot correct for since we do not have E(B-V) values from age-dating in the optical).

We find that the \citet{rodriguez24}  {\em nearly embedded cluster candidates have essentially the same infrared  SEDs as the  1-3~Myr clusters from our optically-identified training set}.{
This is consistent with the results from  \citet{rodriguez24}
who
find that the subsample of 3.3 $\mu$m-enhanced clusters with no HST emission have very similar SED profiles to the subsample of 3.3 $\mu$m-enhanced clusters that do have HST emission.

The \citet{hassani23} sample of nearly embedded clusters shown in the bottom panel looks similar, but has even smaller scatter, probably because the larger apertures (0.67$^{\prime\prime}$ radius)
improve the signal-to-noise, especially for the NIRCAM filters where a smaller 0.124$^{\prime\prime}$ radius aperture is used for the \citet{rodriguez24}
sample. 

The use of different apertures (i.e., ``small aperture'' photometry for \citet{rodriguez24}; large ``convolved'' aperture photometry for \citet{hassani23} - see discussion in Section \ref{sec:small_vs_large}) allows us to test how much effect this has on the resulting SEDs. The NIRCAM  values are nearly identical, but we find a small, systematic change toward shallower slopes for the MIRI filters for the Hassani sample using convolved large aperture photometry, reaching a difference of about a factor of three for the F2100W filter in Figure \ref{fig:embedded}.  This can be understood by the fact that the F2100W observation uses essentially the same aperture, while the F200W aperture has increased a factor of 5.4 in radius, or nearly 30 in area. As can be seen in a variety of images (e.g., Figure \ref{fig:embed_mos_1}), there are typically several other clusters or stars within the 0.67$^{\prime\prime}$ aperture (shown as the yellow dotted line), hence the flux will be considerably larger for the F200W measurement, and hence the normalized value of F2100W / F200W used in the SED will be lower when using the large aperture. A similar result is obtained in Appendix B when comparing the  convolved and small aperture photometry for the training sample.

While the factor of three difference in the top (small aperture photometry) and bottom (convolved aperture photometry) at F2100W is an important component, we note that the overall change in the F2100W flux is a factor of about 100 between a  1 and 5 Myr cluster, and hence is the dominant influence when estimating the ages of the clusters. See Appendix B for a similar comparison using the 40 clusters in the training sample.

\begin{figure}
\begin{center}
\includegraphics[width =3.3in , angle=0] {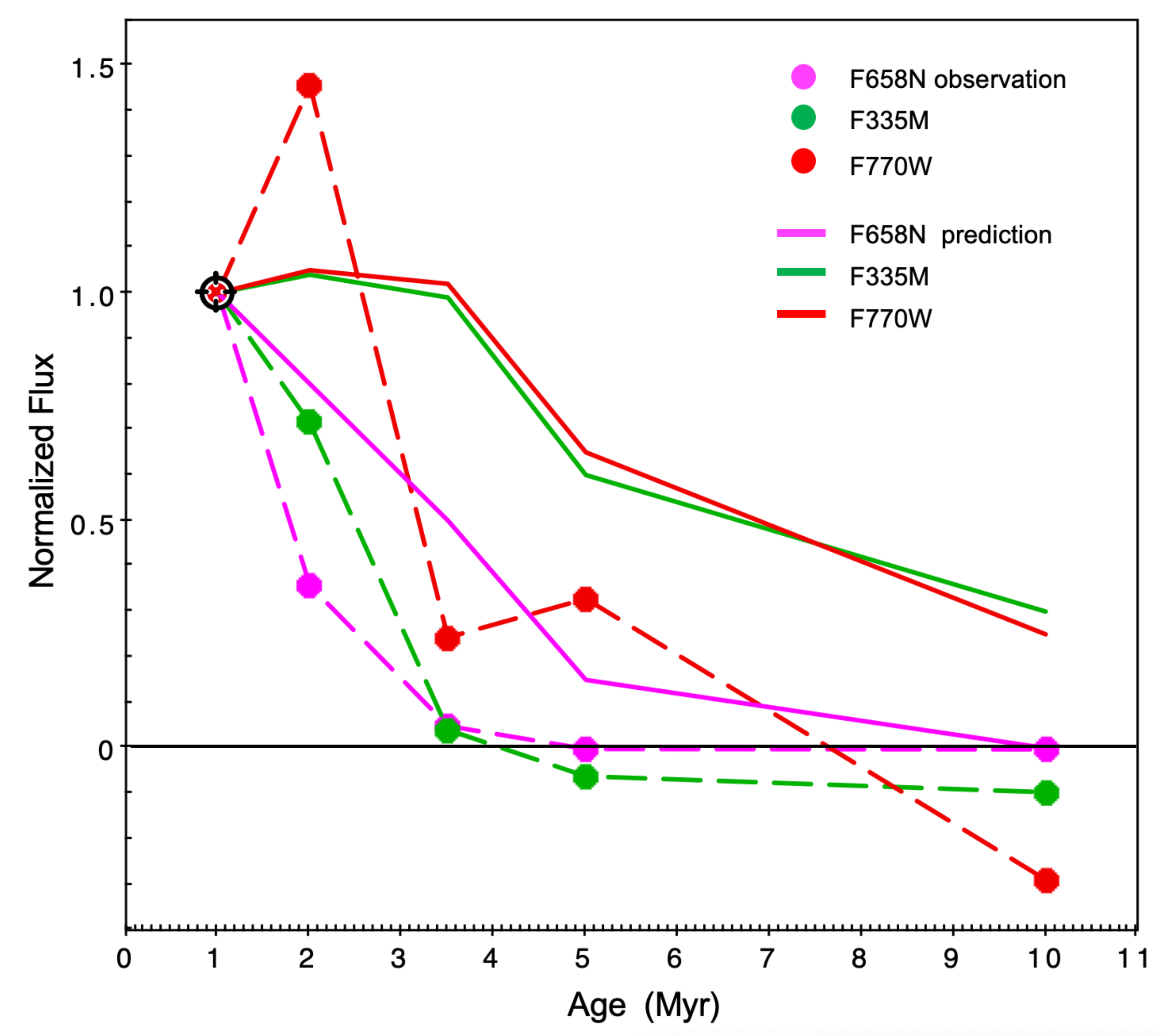}
\
\end{center}
\caption{Observed and predicted evolution of H$_{\alpha}$, F335M and F770W emission. 
The solid lines are from the predictions from the CIGALE/Draine models (see also Figure 7 in \citealt{whitmore23a}), which are static models that do not take into account outflows due to feedback.
% in Figure \ref{fig:cc_draine_original_plot}. 
The dashed lines are from our empirical templates based on the observations, and normalized to 1.0 at an age of 1 Myr. In all cases the  observed profiles are much steeper than the predicted values, and are nearly zero by 4 Myr for the F335M and H$_{\alpha}$ profiles. 
}
\label{fig:temporal_profile}
\end{figure}

\subsection{Temporal Evolution of H$_{\alpha}$,   3.3 $\mu$m,  and 7.7 $\mu$m PAH Emission}
\label{sec:temporal_profiles}

\bigskip

In Figure \ref{fig:temporal_profile} we use our mean-value empirical SEDs to plot the evolution of the 3.3
$\mu$m and 7.7 $\mu$m PAH band strength and compare them with the H$_{\alpha}$ evolution.  
The continuum flux has been estimated and subtracted using the mean of the F555W and F814W flux values for the H$_{\alpha}$ measurement, the F300M flux  for the F335M measurement, and the F1000W flux for  the F770W measurement. The 3 and 4 Myr points have been averaged together to reduce the scatter.
As expected, based on our examination of Figures \ref{fig:1_4_templates}
 and \ref{fig:5_500_templates}, 
 the 3.3 $\mu$m and $7.7 \mu$m PAH lines fall off quickly, and essentially disappear by $\sim5$~Myr.  
  This supports the primary result of this study, that stellar feedback can remove or destroy most of the gas and dust from the small apertures used for our study within just a few Myr. 
 
 We also note a similar evolution for H${\alpha}$ emission.
This new result will be important for including H$_{\alpha}$ in future star cluster age dating studies 
since the presence of H$_{\alpha}$ is generally  considered to be evidence of an age less than 7 Myr (e.g., \citealt{leitherer99}), rather than less than 4 Myr as found in the current study. While there are stars with enough UV flux to ionize the gas and hence produce H$_{\alpha}$ emission for ages around 5 Myr, it appears that the gas has already been removed from the immediate area around the cluster 
by this age, resulting in few or no clusters with  H$_{\alpha}$ emission in the 5 - 7 Myr age range. Hence, in both the H$_{\alpha}$ and PAH emission the dominant physical mechanism responsible for controlling the temporal evolution appears to be the destruction or removal of the gas and dust.

 As noted in Section \ref{sec:training_set},  Figure \ref{fig:1_4_templates} suggests that the 3.3    $\mu$m emission is weaker and more disjointed than H$_{\alpha}$ at a given age in a number of cases.
 Somewhat surprisingly, Figure \ref{fig:temporal_profile} does not appear to be consistent with this interpretation, since the profiles look nearly identical for the two bands. 
 A more careful treatment using a larger sample will be required to make a definitive determination of the relative temporal evolution of  H$_{\alpha}$ and 3.3 $\mu$m in the future.

The longer wavelength PAH features may last longer since they originate from larger dust grains \citep{Baron24}. While Figure \ref{fig:temporal_profile} provides some evidence for this at   7.7 $\mu$m, since there is still weak emission out to 5 Myr, it is difficult to make any conclusions at older ages based on our observations in the 7.7 and 11.3 $\micron$ PAH bands since they reach the detection level at about 5 Myr. 
A few older clusters in Figure  \ref{fig:5_500_templates} appear to show some evidence of associated weak F770W emission (i.e., 3433  in the 5 Myr sample and 1187 and 4356 in the 10 Myr sample), but with such small number statistics this is very uncertain.  Our future study of the larger sample will address this topic more carefully.

\subsection{Color-Color Diagrams}
\label{sec:color-color}

\begin{figure*}
\begin{center}
\includegraphics[width =7in , angle= 0]{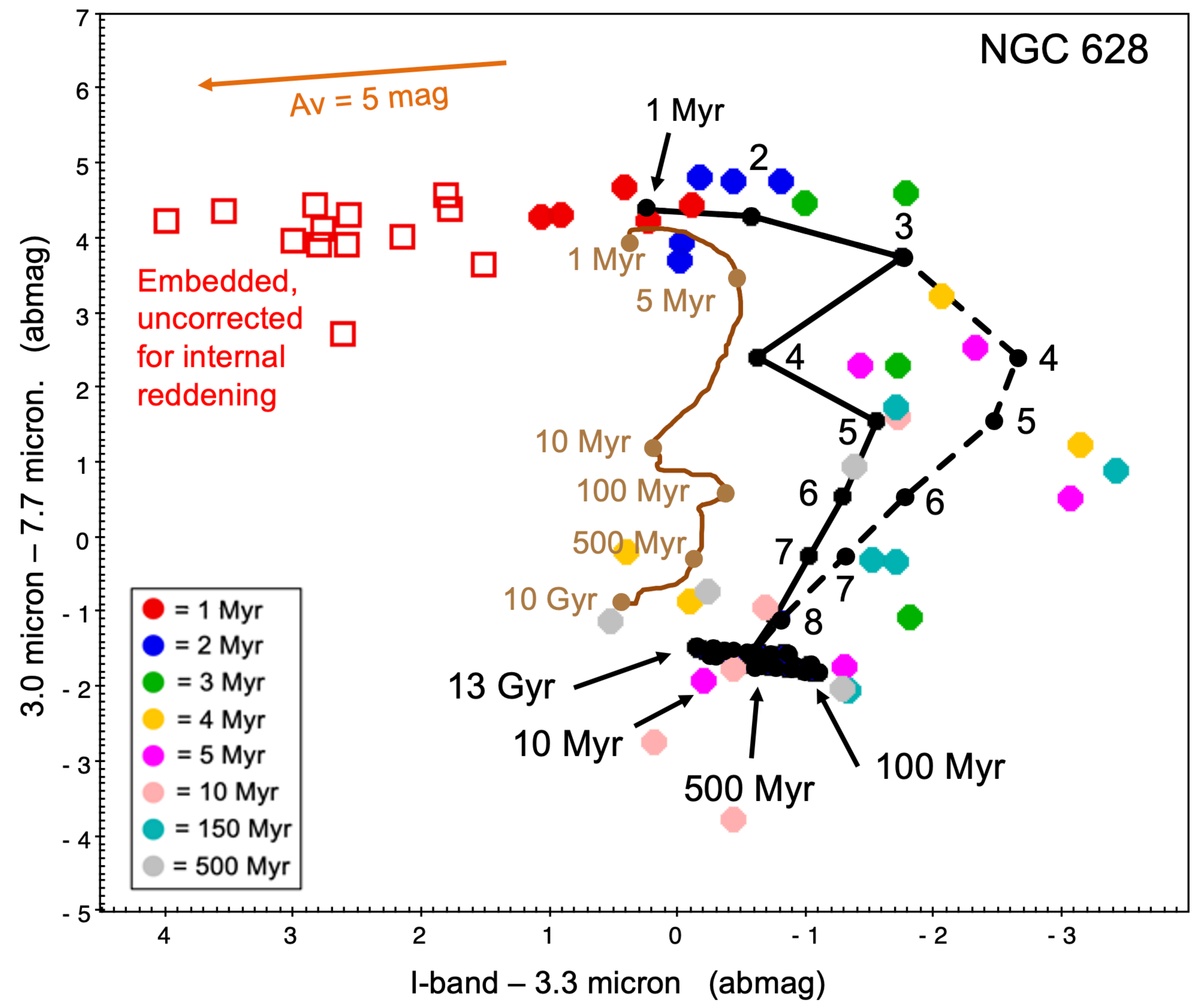}
\
\end{center}
\caption{An I - 3.3 ${\mu}m$ vs 3.0 - 7.7 ${\mu}m$ diagram for our training set (color-coded filled circles).  The  black line is the empirical template based on our training set using means out to 8 Myr, with ages marked by black dots. The dashed black line shows the version with the BC03 stellar continuum used in the near-IR for ages in the 4 - 8 Myr range (i.e., the lower-limit template - see text for discussion).
The brown line is a CIGALE/Draine model from Henny et al. 2025 (submitted), as discussed in the text. 
Note that there is good agreement for the 1 and 2 Myr predictions but a dramatic difference for older clusters. 
The red squares are the nearly embedded  clusters from \citet{rodriguez24} discussed in Section \ref{sec:embedded}. Unlike the other data points, they have not been corrected for reddening. A reddening vector with $A_V = 5$ mag is shown in orange. }
\label{fig:cc_ir}
\end{figure*}

\begin{figure*}
\begin{center}
\includegraphics[width =7in , angle= 0]{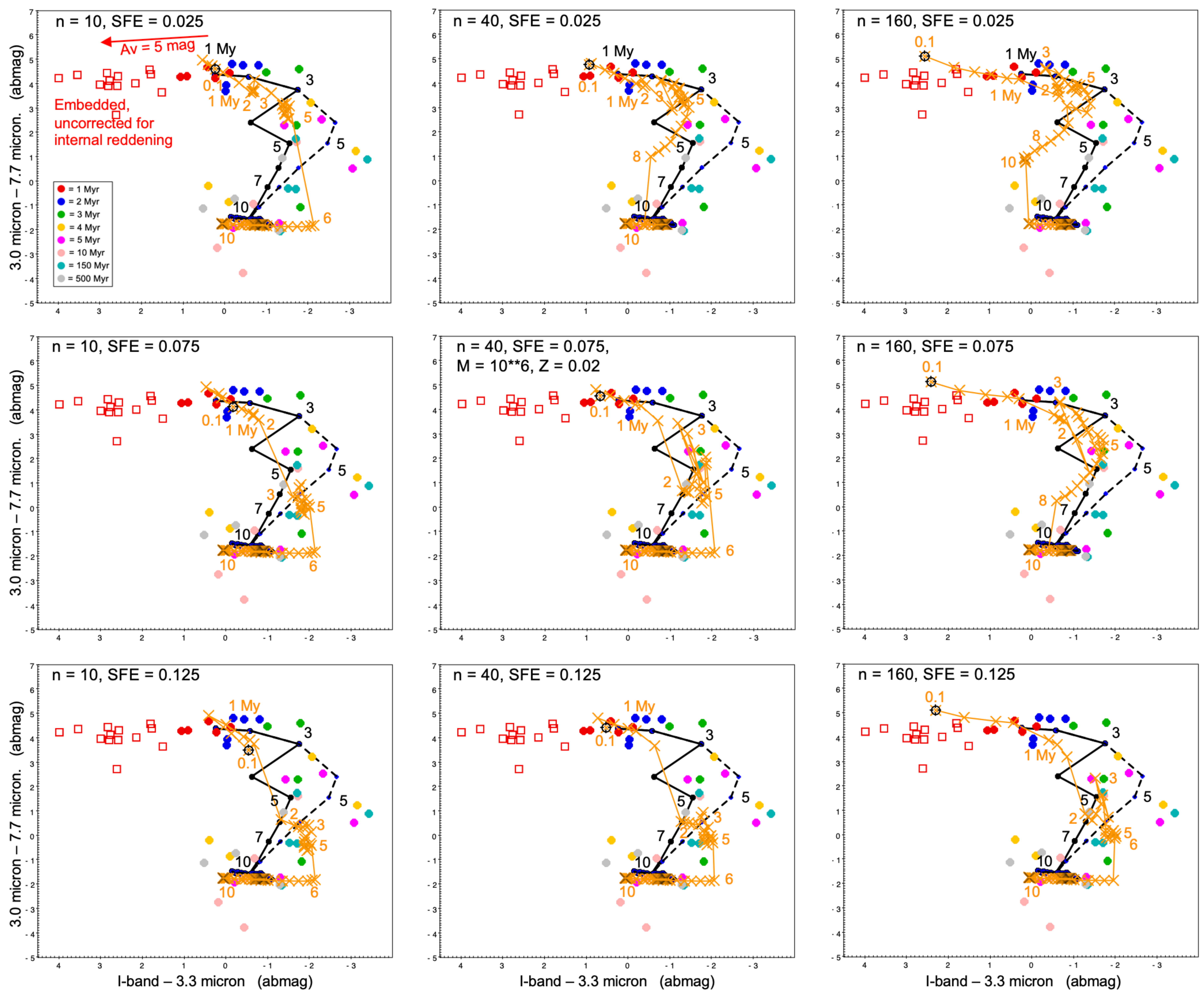}
\
\end{center}
\caption{Similar to Figure \ref{fig:cc_ir}, but overplotting a grid of TODDLERS (\citealt{kapoor23}  - orange crosses and ages)  models covering a range from n = 10, 40, 160 cm$^{-3}$ and SFE = 0.025, 0.075, 0.125  around a central model that shows relatively good agreement with the empirical template models. A model age of
0.1 Myr is shown by  the black open circles. }
\label{fig:cc_toddlers_mos}
\end{figure*}

With initial empirically-based template SEDs in hand, we can now make optical-IR color-color diagrams similar to the UV-optical color-color diagrams that have been such a useful tool in the past. 
While other color combinations are possible, in this paper we only consider the F814W - F335M (i.e., I band - 3.3 $\mu$m) vs. F300M - F770W (i.e., 3.0 $\mu$m - 7.7 $\mu$m) diagram.  Some advantages of this combination are: 1) F814W is the reddest optical band that is still detected for most nearly embedded clusters, hence including this filter provides the widest possible wavelength baseline, 2) the use of continuum  (F814W and F300M) and PAH bands (F335M and F770W) in each axis helps separate the young and old objects, with a spread of $\approx5$ or 6 magnitudes in each color over the age range from 1~Myr through 10~Myr, 3) F770W is the shortest wavelength MIRI band in our dataset and hence has better spatial resolution than the other MIRI filters at longer wavelengths. The ABmag system is used for this figure.

Figure \ref{fig:cc_ir} includes the colors for 37 of the 40 clusters in the training set (i.e., from Table \ref{tab:table_1} - excluding the three clusters discussed as outliers in Section \ref{sec:outliers}) plotted as color-coded filled circles in the figure; the empirical SED templates using mean values derived in this paper (black line and small black dots showing various ages); the empirical SED templates using the BC03 stellar continuum values for the NIRCAM filters for ages greater than 3 Myr (i.e., the lower-limit templates as discussed in Section {\ref{sec:templates});
and the nearly embedded clusters from \citet{rodriguez24} as open red squares. The \citet{hassani23} sample of nearly embedded clusters is not included since it uses a different photometric methodology (i.e., convolved images matched to the 0.67$^{\prime\prime}$ F2100W resolution) and hence cannot be put precisely on this diagram. As expected, the youngest clusters, with strong F335M and F770W emission, tend to show up in the upper left part of Figure \ref{fig:cc_ir} (i.e.,  the red - 1 Myr and blue - 2 Myr circles, and all the embedded objects). The older clusters, e.g., 10 Myr (pink), 150 Myr (cyan), and 500 Myr (grey), tend to be in the bottom center part of the diagram.

We note that the scatter in the observed data points for the 3 (green), 4 (yellow), and 5 (maroon) Myr data points is much larger than the 1 (red), 2 (blue) and 10 Myr (pink) data points. This is most likely due to the very irregular, often patchy bubble morphologies of the 3 through 5 Myr clusters, as seen in Figures \ref{fig:1_4_templates}  and \ref{fig:5_500_templates}. Anticipating the discussion on the TODDLERS models  \citep{kapoor23} in Section \ref{sec:toddlers}, this also may reflect the ``looping'' nature of the colors in this age range. The relatively large difference between the empirical templates using the mean (solid black line) versus BC03 stellar continuum models for 4 - 8 Myr (dashed black lines) also reflects this larger scatter and difference in morphology. A larger training set in the future should help clarify whether this is primarily observational scatter, or an inherent  difficulty in age dating objects in this age range due to differences in morphology.

All of the \citet{rodriguez24} nearly embedded cluster candidates (red squares)  appear to the left of even the youngest clusters in the training set. This is primarily  due to large values of extinction and reddening that has not been corrected since most of these objects do not have age estimates from HST and hence do not have measured E(B-V) values, unlike the other points in the figure.

The brown line in Figure \ref{fig:cc_ir} represents an implementation of the Draine et al. (2021) models in the CIGALE framework
%CIGALE/Draine model 
that Henny et al. 2025 (submitted) found
matched the observational data in NGC 628 relatively well for very young clusters. It has 
A550 = 1.0 mag, Q$_{\rm pah}$ = 0.65, $\alpha$ = 2.0, and f$_{\rm esc}$ = 0.8. 
Note that while there is good agreement between the 1 and 2 Myr predictions and the observed training set points, there is a dramatic difference for older clusters. % as also seen in Figure \ref{fig:cigalle_vs_templates}. 
This discrepancy is probably due to the fact that the ISM is removed or destroyed from the immediate vicinity of a cluster on short time scales, making the assumption of energy-balance used in the current implementation of the 
\citet{draine21} models in the CIGALE framework \citep{boquien19}
inappropriate for older clusters.

\bigskip

\section{Discussion}\label{sec:discussion}

\subsection{Comparison with Predictions from TODDLERS Models}
\label{sec:toddlers}

The TODDLERS (Time evolution of Observables
including Dust Diagnostics and Line Emission from Regions containing young Stars) model SED library developed by \citet{kapoor23} are designed specifically for the dynamic situation around a young star cluster, as they follow a homogeneous gas cloud as it evolves due to stellar feedback processes such as stellar winds, radiation pressure, and supernovae.

The bottom left panel of Figure 19 from \citealt{kapoor23} shows the resulting 
UV-through-mm SED for a model with solar metallicity clouds, number density  n$_{\rm cl}$ = 80 cm$^{-3}$, and a star-forming efficiency of 5 \%. Models with ages 0.1, 0.5, 1, 2, 5, 7, 10 Myr are included. 
The models are in relatively good agreement with our empirical templates shown in our Figure \ref{fig:mean_templates}, with rapidly rising flux values in the mid-IR for the first few Myr, but  much flatter SEDs for ages 5 Myr and older. 

A more detailed comparison can be made from the color-color diagram in Figure \ref{fig:cc_toddlers_mos}, with nine TODDLERS models shown in orange. The models show a number of remarkable similarities with our empirical templates shown in black (solid line for the version using means out to 8 Myr; dashed line for version using the stellar continuum for NIRCAM bands from 4 to 8 Myr), with the large color-coded filled circles showing the training sample, as discussed in Figure \ref{fig:cc_ir}. One of the models that shows the best resemblance is included in the center of the figure. It has  n = 40 cm$^{-3}$, SFE = 0.075, M = 10$^6$ M$_{\odot}$, and z = 0.02 (i.e., solar metallicity). 

A grid of TODDLERS models covering a range from n = 10, 40, 160 cm$^{-3}$ and SFE = 0.025, 0.075, 0.125  around this central model is included using orange crosses and lines, with age estimates from 0.1 (the black open circle) to 10 Myr labeled in orange.

Several TODDLERS models show good agreement with the observations and empirical templates in three respects:
%% \begin{itemize}
1) the position of the 1 Myr clusters in F814W - F335M vs F300M - F770W color-color space, 2) the positions of the 2 to 5 Myr clusters, although some show a roughly linear progression while others show a looping (in color-color space) evolution in this age range, 
3) the position of clusters older than 10 Myr (i.e., as predicted by the BC03 stellar continuum models). 

%% \end{itemize}

We also note that several aspects of the models vary dramatically, providing good diagnostic power. These include: 
%%\begin{itemize}
1)    the youngest (0.1 Myr) models have F335M emission that is  much too strong (left) in F814W - F335M for all the n = 160 cm$^{-3}$ models, 
2) the large gaps (rapid evolution) between certain young and old ages  (e.g., between 1.7 and 2 Myr in the central panel),
3) the location of the intermediate age (6 to 8 Myr) clusters swings rapidly from F814W - F335M = -2.0 to 0.0  as a function of SFE for the n = 40 and 160 cm$^{-3}$ models.

While the good correspondence between some of the TODDLERS models and the empirical templates is  encouraging, it is not the main focus of this pilot study, hence we leave a more detailed comparison for the future when more accurate empirical templates are available.
We are also investigating the incorporation of the TODDLERS model SED library, as well as some aspects of the empirical templates discussed in the current paper, into CIGALE  
\citep[Code Investigation Galaxy Emission]{boquien19} in the future.

\subsection{Comparison with Related Observational Work}
\label{sec:previous}

One of the primary results from our study is that only star clusters with ages less than about 5 Myr have strong PAH or infrared dust continuum emission. A similar result has also been reported in \citet{rodriguez23} based on selection of cluster candidates using the 3.3 $\micron$ image. They found  that most of the 3.3 $\mu$m emitters  have ages less than 2 Myr. 

The slightly older limiting age found in the current study may be 
% largely 
due to the use of SEDs that include MIRI observations out to 21 $\mu$m, and hence may reflect  the slower destruction  of the larger dust grains responsible for the 7.7 through 21 $\mu$m emission compared to the 3.3 $\mu$m emission \citep{sandstrom23a,EGOROV_PHANGSJWST,chastenet23}. However, the \citet{rodriguez23} result might also be a slight underestimate due to the 1 Myr bias in the \citet{turner21} ages, as discussed in \citet{rodriguez24} and Thilker et al. 2025 (submitted). The more recent \citet{rodriguez24} paper finds that most of the 3.3 $\mu$m emitters  have ages less than  about 3 Myr.

A related result is the finding that PAH emission is suppressed in HII regions \citep{churchwell06,relano09,sandstrom23a, EGOROV_PHANGSJWST,chastenet23}. It seems likely that both our result that most clusters with ages greater than 5 Myr do not have strong PAH or IR dust continuum emission, and the results on suppressed PAH emission in HII regions, are caused by a combination of the same two physical mechanisms, i.e.: 1) the removal of gas and dust in the central region of superbubbles due to outflows from the star cluster, and 2) the destruction of dust grains resulting in the reduction of PAH emission. 
The relative importance of the two effects may depend on what size aperture is used for the observations.

In the current study the focus is on determining ages of the star clusters, rather than investigating the nature of the HII regions around the clusters. There are relatively few studies that have examined PAH and IR emission from the star clusters themselves in external galaxies. However, two studies based on Spitzer observations suggest that PAH emission may be associated with clusters with ages in the tens or even hundreds of Myr range \citep{mallory22,lin20}. This would be in contradiction to the results from the current paper. It is possible that some of these cases are superpositions, since the initial selection of the sources in these studies is based on detection of IR emission in Spitzer observations, and the subsequent matching with the star cluster is based on position within the relatively large Spitzer PSF. JWST observations of these objects should provide a more definitive determination of whether the IR emission is actually associated with the older clusters or is randomly situated in these cases.   

A number of other studies have also established a link between older stellar populations and PAH emission, including \citet{draine21} - Figure 16, and \citet{leroy23}. It is important to note that these results are for integrated stellar populations over a relatively large field of view, for example parts of the bulge in M31 in the case of \citet{draine21}, or the diffuse dust lanes in PHANGS galaxies in the case of \citet{leroy23}. The results  from the current study are primarily relevant for cases where individual star clusters have been observed using small apertures.

A recent study by \cite{pedrini24}, taken as part of the 
FEAST project (Feedback in Emerging Extragalactic Star Clusters, Adamo et al. 2025 - in prep), and also based on HST$+$JWST observations of NGC 628, 
finds several similar results and hence supports some of the basic conclusions from the current paper.  There are, however, some important differences.
\cite{pedrini24} find that strong PAH emission is only found in clusters younger than 7 Myr, with the most common ages being in the  3 - 6 Myr range. In the current paper we find that strong PAH emission is only found for clusters less than about 5 Myr old, with the most common emission in the 1  - 2 Myr range.  
While the details of the age dating are not yet available for the FEAST study, 
%(i.e., Linden et al. 2024 - in prep), 
it is likely that the primary difference is due to our use of a new and largely independent age dating approach, as described in Section~2.

Another result of the \cite{pedrini24} study is the finding 
that the 3.3 $\mu$m and 7.7 $\mu$m bands have very similar temporal profiles (their Figure 9), while we find that the 3.3 $\mu$m band declines faster than the 7.7 $\mu$m profile, as shown Figure \ref{fig:temporal_profile}.
While this remains an open question, a variety of recent observational results \citep{EGOROV_PHANGSJWST,chastenet23, Baron24} suggest fast destruction of the smaller grains responsible for the 3.3 $\mu$m PAH emission is commonly found.

\section{Summary and Conclusions}
\label{sec:conclusions} 

In this pilot study we use new infrared PHANGS-JWST NIRCAM and MIRI imaging of the spiral galaxy NGC~628 in the F200W, F300M, F335M, F360M, F770W, F1000W, F1130W, and F2100W filters, in addition to existing HST observations in the F275W, F336W, F438W, F555W, F658N, F814W filters, to  
% to illustrate a number of points and to 
produce empirical SED templates for star clusters. These templates can be used both to provide age estimates for star clusters, and to test new models such as the TODDLERS (Time evolution of Observables
including Dust Diagnostics and Line Emission from Regions containing young Stars) SED library developed by \citet{kapoor23}. Our primary conclusions are provided below.

1. Nearly all star clusters with strong PAH and IR dust continuum emission have ages in the 1 – 4 Myr age range (also see \citealt{rodriguez23} and \citet{rodriguez24}. 
In nearly all cases where there is strong PAH emission there is also strong H$\alpha$ emission.

2. In this pilot study we develop %tentative 
empirical
%ly determined  
 SED templates based on a carefully chosen training set of 40 %training 
clusters in NGC 628, ranging in age from 1 to 500 Myr, with masses greater than $\approx3000~M_{\odot}$. When combined with stellar continuum models from BC03 models, which agree with observations of all clusters with ages greater than 10 Myr, we are able to provide 
% empirical 
SED templates for clusters of all ages.   

3. Using these templates we plot the observed evolution of H$\alpha$ and PAH (3.3 $\mu$m and 7.7$\mu$m) strength and demonstrate that they are similar and all decrease rapidly (within a few Myr), but with the PAH 3.3 $\mu$m emission  dropping faster than the 7.7 $\mu$m emission and possibly faster than H$\alpha$. These differences are likely influenced by the rapid destruction of  small dust grains responsible for the 3.3 $\mu$m emission.  

4. The rapid decline of PAH and IR dust continuum emission with age is probably due to a combination of the destruction of the dust grains and stellar feedback removing the gas and dust from the immediate vicinity of the star cluster. The time scale for the latter mechanism is compatible with dynamical estimates based on CO superbubble expansion velocities from \citet{watkins23} in NGC 628.

5. Samples of nearly embedded cluster candidates in NGC 628 from \citet{hassani23} and \citet{rodriguez24}  are examined and found to have SEDs and colors very similar to the 1 to 3 Myr clusters from the optical training set in the IR part of the spectrum. The Hassani sample shows less scatter, and is most consistent with the 1 Myr convolved aperture template.   In nearly all cases we find there is a young star cluster within a few tenths of an arcsec (10 - 30 pc) of the nearly embedded cluster in the Hassani sample, suggesting  the formation of the embedded object was triggered by its presence.

6. Using these empirical templates we produce   evolutionary tracks for the  0.8 - 3.3 $\mu$m  vs 3.0 - 7.7 $\mu$m color - color diagram that are in agreement with the optically selected training set, the samples of nearly embedded cluster candidates, and the current implementation of the Draine models within CIGALE for ages of 1 to 2 Myr. For ages older than about 5 Myr the CIGALE/Draine models overpredict the strength of the PAH and thermal dust emission, probably due to the assumption of energy balance.

7. We find that the empirical SED templates are in fairly good agreement with the TODDLERS \citep{kapoor23} model SED library which is based on the dynamic, spherical evolution of a homogeneous gas cloud around a young stellar cluster. The model SEDs support our results that PAH and dust continuum emission dominates in the near-IR for  only the first few Myr, and emission in the Mid-IR is dominated by the stellar continuum after roughly 5  Myr.

In this pilot study we  report on results for a single galaxy (NGC~628 which is located at 9.84~Mpc), and a specific set of small apertures used for the HST and JWST photometry. We briefly examine the effects of using images convolved to the size of the F2100W point spread function (0.67$^{\prime\prime}$) and find that it is an important, but not dominant effect. We caution the reader that our quantitative results depend at least somewhat on these specific properties and assumptions, although we believe that the main conclusions are robust regardless of the specific details. 
These issues will be explored more fully in the future, when a much larger cluster sample will be used to determine higher quality empirical SED templates for star clusters.

\section{Acknowledgements}

%%%\begin{acknowledgements}

This work is based on observations made with the NASA/ESA/CSA James Webb Space Telescope (program \#2107) and the NASA/ESA Hubble Space Telescope (program \#15654 \& \#13364).   
The data were obtained from the Mikulski Archive for Space Telescopes at the Space Telescope Science Institute, which is operated by the Association of Universities for Research in Astronomy, Inc., under NASA contract NAS 5-03127 for JWST and 5-26555 for HST.

%\vspace{5mm}
\facilities{{\em HST} (Hubble Space Telescope), {\em JWST}, ALMA (Atacama Large Millimeter/submillimeter Array),  VLT-MUSE (Very Large Telescope - Multi Unit Spectroscopic Explorer).}\\

\software{ASTRODENDRO, AstroPy, CIGALE, Photutils, TOPCAT.} \\

We thank the referee for a careful reading of the paper and several suggestions that greatly improved the paper.

AW acknowledges UNAM's DGAPA for the support in carrying out her sabbatical stay at UCSD, USA, through program PASPA.

KG is supported by the Australian Research Council through the Discovery Early Career Researcher Award (DECRA) Fellowship (project number DE220100766) funded by the Australian Government. 
KG is supported by the Australian Research Council Centre of Excellence for All Sky Astrophysics in 3 Dimensions (ASTRO~3D), through project number CE170100013.

MB gratefully acknowledges support from the ANID BASAL project FB210003 and from the FONDECYT regular grant 1211000.
This work was supported by the French government through the France 2030 investment plan managed by the National Research Agency (ANR), as part of the Initiative of Excellence of Université Côte d’Azur under reference number ANR-15-IDEX-01.

MC gratefully acknowledges funding from the DFG through an Emmy Noether Research Group (grant number CH2137/1-1). COOL Research DAO is a Decentralized Autonomous Organization supporting research in astrophysics aimed at uncovering our cosmic origins.

\section*{Data Availability}

The imaging observations underlying this article can be retrieved from the Mikulski Archive for Space Telescopes at \url{https://archive.stsci.edu/hst/search_retrieve.html} under proposal GO-15654. High level science products, including science ready mosaicked imaging, associated with HST GO-15654 are provided at \url{https://archive.stsci.edu/hlsp/phangs-hst}.
The specific PHANGS-JWST observations analyzed can be accessed via \dataset[10.17909/9bdf-jn24]{http://dx.doi.org/10.17909/9bdf-jn24} and PHANGS-HST images accessed via \dataset[10.17909/t9-r08f-dq31]{https://dx.doi.org/10.17909/t9-r08f-dq31}. 
The full version of the empirical templates (both the mean-value and the lower-limit versions) can be accessed via 
\doi{10.17909/jray-9798}.
% doi:10.17909/jray-9798 .

        }

\bibliography{all_nov_24_2021,phangsjwst}{}
\bibliographystyle{aasjournal}

%%% \clearpage

\appendix

\section{A Closer Look at the 1 - 4 Myr Box}
\label{append:1-4box}

Figure \ref{fig:1-4box_mosaic} provides a more detailed look at the 45 (of 58) clusters within the 1 - 4 My box that fall within the field of view of the F770W image. Thirty one of these 45 clusters have strong PAH emission, defined as F770W / F300M flux $>$ 20 , as in Figure \ref{fig:cc_single_age}. This figure supports several assumptions discussed in Section \ref{sec:non-degenerate}, and provides additional insight into the youngest clusters.

The figure is broken into six classifications, which are shown for the strong PAH sample on the left and the weak PAH sample on the right. The sequence is roughly in order of age, with compact H$_{\alpha}$ objects on the left, followed by small superbubbles, clusters near the edge of the shell of superbubbles, and large superbubbles in the first four columns.
Clusters which are more uncertain (e.g., possible superpositions) are in the fifth column, followed by clusters with no obvious  H$_{\alpha}$ (just 2 of the 45) in the last column.

Clusters without O and B stars that are not massive enough (hot enough) to ionize hydrogen atoms are not expected to have H$_{\alpha}$ emission, and this is apparently the case for just two clusters (2446 and 4275). 
In two more cases (4460 and 5030) there is H$_{\alpha}$ in the area, but it is not clear whether it is associated with the cluster (i.e., no clear bubble or morphological shape that suggests a connection).
This suggests that in 27 / 31 = 87 \% of the strong PAH sources, there is H$_{\alpha}$ associated with the clusters, in agreement with our discussion of Figures \ref{fig:image_and_ages_halpha} and \ref{fig:image_and_ages_770} in Section \ref{sec:observations}.

In general, the large bubbles  have weak PAH emission, since most of the gas and dust has been blown outside the effective aperture used for our study (i.e., the orange dashed circle for the F770W  and the dotted yellow circle for the F2100W effective aperture). We note that most of the weak PAH sources on the right side of Figure \ref{fig:1-4box_mosaic} also appear to have some H$_{\alpha}$ associated with them. Hence the vast majority of all the clusters in the 1 - 4 My box are very young, as predicted based on the discussion of non-degenerate regions in Section \ref{sec:non-degenerate}.
Finally, we note that there is a roughly even distribution of clusters from very compact ($\approx$ 1 Myr) to large bubbles and uncertain objects ($\approx$ 4 Myr), which was one of the assumptions used in Section \ref{sec:bubbles} .

\begin{figure*}
\begin{center}
\includegraphics[width =7in, angle= 0]{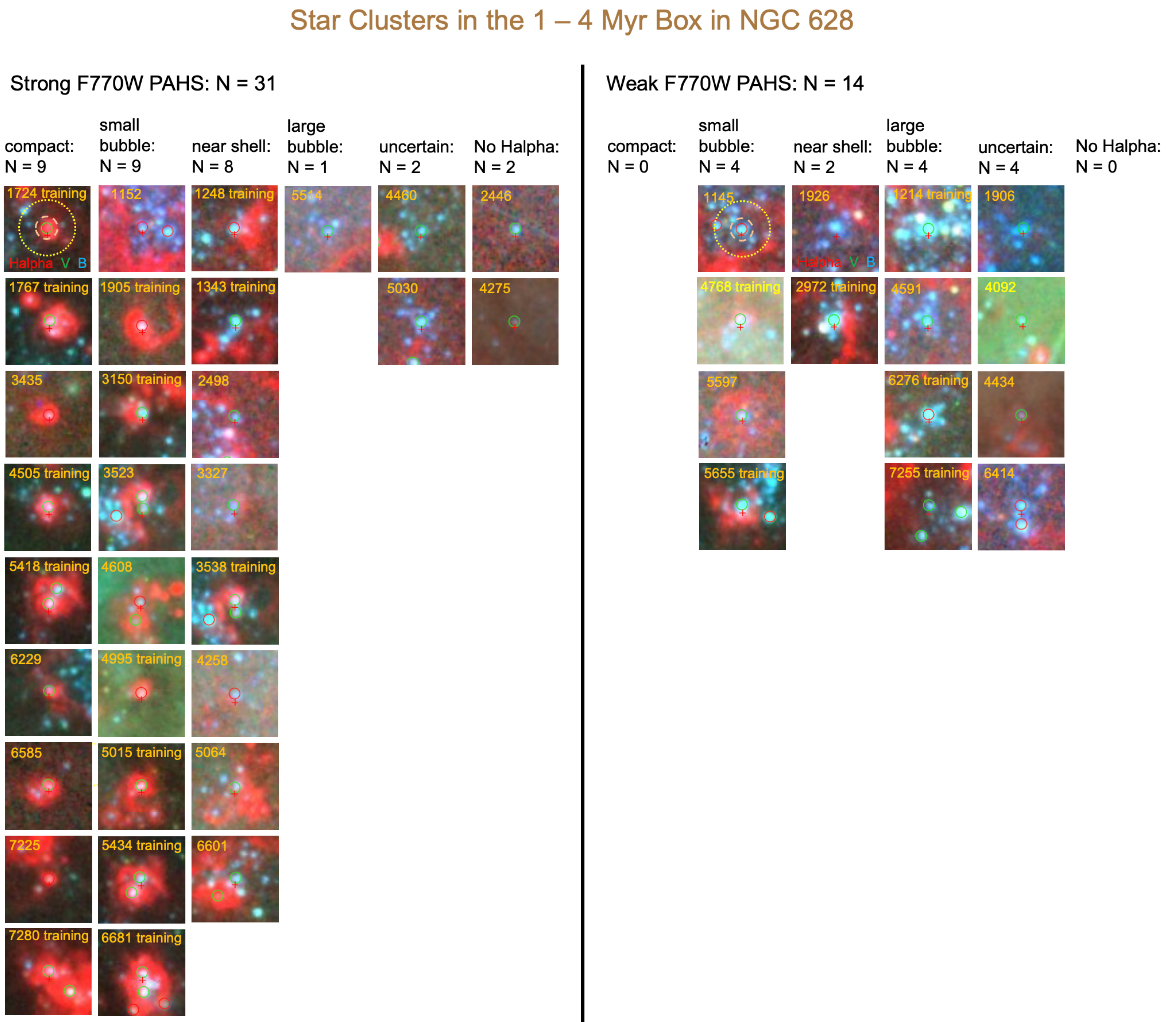}
% {pilot2_5_boxes_mar_13_2024.pdf}
\end{center}
\caption{ H$_{\alpha}$, V, B images of the  45 sources within the F770W FOV in the 1 - 4 My Box. The 20 clusters included in the training set (i.e., highest mass objects)  are also identified in the snapshots. Red circles are used for class 1 clusters while green circles are used for class 2 clusters. Aperture sizes, as included in Figure {\ref{fig:region_1}, are shown in the first panel.}
}
\label{fig:1-4box_mosaic}
\end{figure*}

\clearpage

\section{Convolved Large Aperture Photometry of the Training Set }
\label{append:convolved_app}

In Section \ref{sec:small_vs_large} and \ref{sec:embedded} we briefly discussed the use of convolved large aperture photometry. In this appendix, we take a closer look at SEDs derived using convolved aperture photometry from \citet{hassani23} of clusters in our optically-selected training set, and comparee with SEDs derived using the small-aperture photometry discussed in Section \ref{sec:small_vs_large}. The photometry used in this exercise is included in Tables \ref{tab:table_7} and \ref{tab:table_8}.

Figure \ref{fig:convolved_seds} shows a blowup of part of the bottom panel from Figure \ref{fig:embedded}, this time including SEDs derived from photometry based on convolved
images from \citet{hassani23} (i.e., the dashed color-coded lines). The first thing to notice is that the observations for the 12 nearly embedded cluster candidates discussed in Section \ref{sec:embedded} (the brown points) agree best with the SEDS derived from the convolved images for the 1 Myr clusters (i.e., the red dashed line). This is similar to the conclusion for the \citet{rodriguez24} nearly embedded cluster candidates using small aperture photometry, although in that case (upper panel of Figure \ref{fig:embedded}), there was also reasonable agreement with the 2 and 3 Myr templates for a few of the embedded clusters. 

The next point to notice is that the convolved SED templates are systematically shallower than for the small aperture templates for 1 - 3 Myr.  This is  to be expected since the aperture used for the F2100W photometry is unchanged, but the size of the shorter wavelength apertures are dramatically increased (e.g., from 0.12$^{\prime\prime}$ to 0.67$^{\prime\prime}$ for NIRCAM) in the case of the convolved apertures. Hence they include surrounding clusters in many cases, thus increasing the F200W flux and reducing the ratio between F2100W and F200W (e.g., examine the apertures overplotted in the first panels of Figures \ref{fig:embed_mos_1} and \ref{fig:embed_mos_2}). The case is less clear for 4 Myr in Figure \ref{fig:convolved_seds}, probably due to the increased noise from the patchy and erratic bubble structure.
For ages of 5 Myr and older, the MIRI flux is essentially noise, as was the case in Figure \ref{fig:mean_templates}.

\begin{figure*}
\begin{center}
\includegraphics[width =6in, angle= 0]{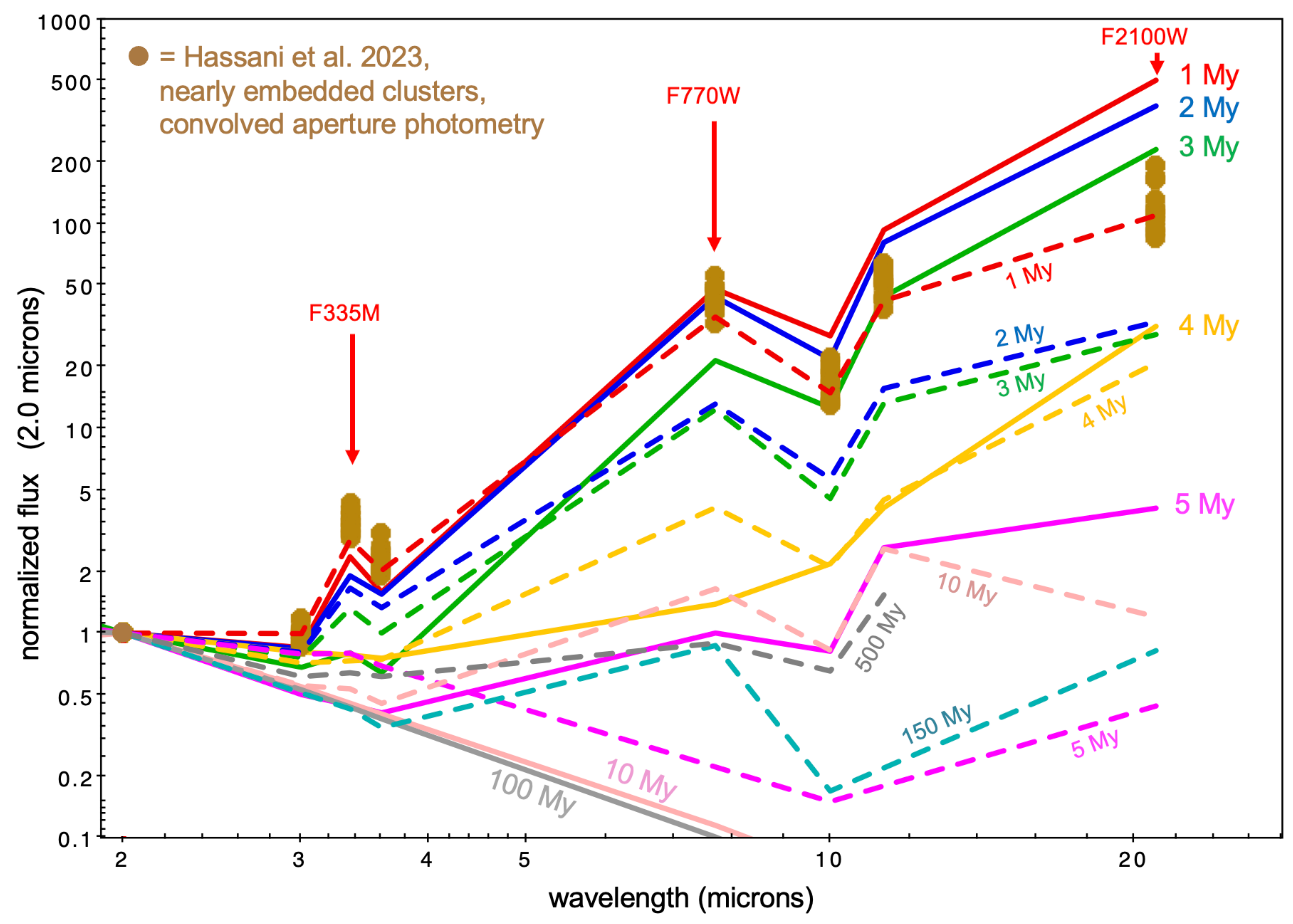}
% {pilot2_5_boxes_mar_13_2024.pdf}
\end{center}
\caption{ {\bf Similar to the bottom panel of Figure \ref{fig:embedded}, but with templates derived from aperture  photometry using images convolved to the size of the F2100W aperture (i.e., 0.67$^{\prime\prime}$) included as color-coded dashed lines. The brown points are from 12 nearly embedded cluster candidates from \citet{hassani23}, as discussed in Section \ref{sec:embedded}, and are most consistent with the 1 Myr convolved aperture template (i.e., the red dashed line). Hence the  nearly embedded cluster candidates from the Hassani sample appear to be slightly younger than most of our  optically selected clusters. }
}
\label{fig:convolved_seds}
\end{figure*}

\begin{table*}
\caption{Normalized Flux Values Using Small Aperture Photometry    for the Training Set$^a$ }
\label{tab:table_7}
 
     \centering

\begin{tabular}{rrrrrrrrrrrrrrrrrr}
  \hline
ID$^b$ & Age$^c$ & 0.275 & 0.336 & 0.435 & 0.555  & 0.658 & 0.814 & 2.0 &  3.0  & 3.3 & 3.6 & 7.7 & 10.0 & 11.3 & 21.0 & 0.814-flux$^d$  \\

\hline  
 
  1724 & 1.0 & 5.15 & 4.07 & 2.38 & 1.9 & 26.48 & 1.0 & 0.7 & 0.55 & 1.78 & 1.1 & 27.79 & 12.5 & 40.7 & 84.4 & 2.86 \\
  1767 & 1.0 & 4.08 & 3.27 & 2.06 & 1.5 & 20.35 & 1.0 & 0.72 & 0.53 & 1.64 & 1.08 & 26.78 & 14.9 & 49.0 & 224.1 & 14.10\\
  4505 & 1.0 & 3.95 & 2.93 & 2.12 & 1.38 & 11.46 & 1.0 & 0.4 & 0.28 & 0.77 & 0.5 & 16.52 & 7.8 & 30.3 & 111.0 & 6.63\\
  5418 & 1.0 & 4.88 & 3.79 & 2.39 & 1.83 & 11.6 & 1.0 & 0.49 & 0.35 & 0.92 & 0.61 & 17.15 & 8.3 & 29.6 & 166.9 & 6.68\\
  7280 & 1.0 & 4.72 & 3.41 & 2.24 & 1.67 & 11.2 & 1.0 & 0.32 & 0.53 & 1.16 & 0.93 & 38.91 & 31.6 & 97.3 & 746.1 & 4.65\\
  3150 & 2.0 & 4.44 & 3.52 & 2.47 & 1.73 & 2.79 & 1.0 & 0.18 & 0.14 & 0.33 & 0.22 & 11.15 & 4.6 & 21.2 & 85.1 & 14.11 \\
  3538 & 2.0 & 4.58 & 3.59 & 2.46 & 1.75 & 8.29 & 1.0 & 0.48 & 0.41 & 0.74 & 0.56 & 14.98 & 7.6 & 28.0 & 169.1 & 8.00\\
  4995 & 2.0 & 4.18 & 2.89 & 2.09 & 1.56 & 8.56 & 1.0 & 0.28 & 0.16 & 0.62 & 0.45 & 13.65 & 6.0 & 21.6 & 80.7 & 3.66\\
  5015 & 2.0 & 4.14 & 3.33 & 2.41 & 1.62 & 8.16 & 1.0 & 0.36 & 0.37 & 0.72 & 0.8 & 10.79 & 6.6 & 17.8 & 40.1 & 4.16\\
  6681 & 2.0 & 4.33 & 3.29 & 2.34 & 1.59 & 5.24 & 1.0 & 0.28 & 0.22 & 0.53 & 0.37 & 17.38 & 9.1 & 36.8 & 207.7 & 8.22\\
  1905 & 3.0 & 4.18 & 3.18 & 2.4 & 1.57 & 4.66 & 1.0 & 0.24 & 0.18 & 0.34 & 0.25 & 10.95 & 4.9 & 18.9 & 43.6 & 3.74\\
  1343 & 3.0 & 4.34 & 3.14 & 2.2 & 1.51 & 1.26 & 1.0 & 0.44 & 0.19 & 0.14 & 0.12 & -0.21 & 0.0 & -0.4 & 1.4 & 12.5 \\
  2972 & 3.0 & 4.29 & 3.35 & 2.47 & 1.69 & 1.59 & 1.0 & 0.19 & 0.17 & 0.18 & 0.16 & 1.37 & 1.6 & 3.6 & 8.1 & 24.0 \\
  4768 & 3.1 & 4.33 & 3.16 & 2.04 & 1.54 & 2.48 & 1.0 & 0.81 & 0.3 & 0.31 & 0.28 & 0.29 & -0.1 & -0.1 & 7.4 & 7.41\\
  5434 & 3.0 & 4.79 & 3.69 & 2.46 & 1.84 & 3.44 & 1.0 & 0.13 & 0.14 & 0.15 & 0.12 & 9.34 & 6.0 & 21.9 & 177.3 & 5.85\\
  1214 & 4.0 & 1.78 & 1.66 & 1.42 & 1.07 & 1.06 & 1.0 & 1.4 & 1.42 & 1.43 & 1.41 & 1.18 & 3.4 & 3.3 & 4.6 & 31.2 \\
  1248 & 4.0 & 3.89 & 3.42 & 2.47 & 1.73 & 1.76 & 1.0 & 0.17 & 0.07 & 0.05 & 0.05 & 0.09 & 0.3 & 1.4 & 16.2 & 5.76\\
  5655 & 4.0 & 4.8 & 3.91 & 2.64 & 1.88 & 1.0 & 1.0 & 0.19 & 0.13 & 0.12 & 0.1 & 2.47 & 1.7 & 6.4 & 71.2 & 13.8 \\
  6276 & 4.0 & 4.22 & 3.24 & 2.28 & 1.48 & 1.32 & 1.0 & 1.17 & 0.73 & 0.68 & 0.63 & 0.32 & 0.9 & 0.8 & -0.1 & 32.5 \\
  7255 & 4.1 & 4.11 & 3.11 & 2.02 & 1.38 & 0.87 & 1.0 & 0.63 & 0.22 & -0.07 & 0.03 & -7.11 & -3.1 & -15.4 & -6.9 & 3.45\\
  1667 & 5.0 & 4.05 & 3.29 & 2.65 & 1.75 & 1.06 & 1.0 & 0.22 & 0.1 & 0.08 & 0.08 & 1.01 & 0.4 & 2.7 & 1.3 & 3.68\\
  2416 & 5.0 & 1.45 & 1.42 & 1.44 & 1.15 & 1.03 & 1.0 & 1.36 & 0.73 & 0.65 & 0.6 & 0.12 & 0.7 & 0.5 & 0.5 & 19.7 \\
  3433 & 5.0 & 2.05 & 2.41 & 2.47 & 1.75 & 1.79 & 1.0 & 0.3 & 0.17 & 0.19 & 0.17 & 1.38 & 0.8 & 3.0 & 8.1 & 6.26\\
  5894 & 5.0 & 1.52 & 1.31 & 1.53 & 1.22 & 1.04 & 1.0 & 0.67 & 0.3 & 0.25 & 0.22 & 0.06 & 0.2 & 0.2 & -0.1 & 10.8 \\
  6895 & 5.0 & 2.93 & 3.04 & 2.81 & 1.98 & 1.37 & 1.0 & 0.12 & 0.05 & 0.03 & 0.02 & 0.14 & 0.1 & 0.7 & 1.3 & 5.47\\
  1187 & 10.0 & .88 & 1.23 & 1.1 & 0.96 & 0.99 & 1.0 & 0.91 & 0.4 & 0.32 & 0.27 & 0.16 & 0.0 & 0.2 & 0.1 & 12.0 \\
  2688 & 10.0 & 1.02 & 0.83 & 0.8 & 0.68 & 0.81 & 1.0 & 1.37 & 0.64 & 0.54 & 0.48 & 0.13 & 0.2 & 0.1 & 0.1 & 25.6 \\
  4356 & 10.0 & 0.94 & 0.79 & 0.67 & 0.58 & 0.49 & 1.0 & 2.02 & 1.05 & 0.89 & 0.8 & 0.18 & 0.7 & 0.5 & 0.7 & 9.61\\
  5016 & 10.0 & 1.06 & 0.84 & 0.82 & 0.67 & 0.74 & 1.0 & 1.41 & 0.63 & 0.52 & 0.46 & 0.04 & 0.2 & -0.1 & -0.4 & 16.4 \\
  7150 & 10.0 & 1.11 & 1.08 & 1.03 & 1.06 & 1.26 & 1.0 & 0.35 & 0.16 & 0.14 & 0.12 & 0.68 & 0.4 & 1.8 & 1.4 & 10.2 \\
  1743 & 150.0 & 0.38 & 0.5 & 1.07 & 0.97 & 1.09 & 1.0 & 0.76 & 0.28 & 0.2 & 0.17 & 0.74 & 0.4 & 1.5 & 9.2 & 3.35\\
  2535 & 150.0 & 0.29 & 0.45 & 0.98 & 0.99 & 1.01 & 1.0 & 0.61 & 0.27 & 0.2 & 0.19 & 1.31 & 0.7 & 2.4 & 1.0 & 2.80\\
  4901 & 150.0 & 0.25 & 0.48 & 0.99 & 1.0 & 1.05 & 1.0 & 0.74 & 0.29 & 0.22 & 0.2 & -1.43 & -0.6 & -3.2 & -0.9 & 2.86\\
  5736 & 150.0 & 0.23 & 0.38 & 0.91 & 0.89 & 0.94 & 1.0 & 0.61 & 0.3 & 0.2 & 0.17 & -0.6 & -0.7 & -3.5 & -3.6 & 2.74\\
  6400 & 150.0 & 0.33 & 0.41 & 0.99 & 0.95 & 0.97 & 1.0 & 0.76 & 0.35 & 0.28 & 0.25 & -0.09 & 0.0 & -0.3 & 0.4 & 7.15 \\
  1516 & 500.0 & 0.16 & 0.23 & 0.83 & 0.84 & 0.85 & 1.0 & 2.77 & 1.41 & 1.46 & 1.43 & 0.49 & 0.6 & 0.7 & 0.2 & 3.09\\
  2273 & 500.0 & 0.13 & 0.22 & 0.8 & 0.86 & 0.75 & 1.0 & 0.63 & 0.3 & 0.25 & 0.23 & 0.72 & 0.7 & 2.4 & 1.6 & 7.14\\
  2352 & 500.1 & 0.06 & 0.2 & 0.67 & 0.94 & 2.54 & 1.0 & 0.63 & 0.35 & 0.46 & 0.35 & 8.3 & 3.9 & 16.7 & 45.7 & 8.98\\
  2984 & 500.0 & 0.01 & 0.16 & 0.56 & 0.82 & 0.77 & 1.0 & 0.72 & 0.3 & 0.25 & 0.23 & -0.06 & 0.2 & -0.0 & -1.0 & 3.75\\
  6340 & 500.0 & 0.01 & 0.2 & 0.67 & 0.77 & 0.91 & 1.0 & 1.7 & 0.75 & 0.72 & 0.65 & 0.38 & 0.1 & 0.8 & 0.7 & 4.02\\

  \hline

 \hline
 
\end{tabular}

 \raggedright

\bigskip

$^a$ Small aperture photometry from \citet{rodriguez24} for clusters in the training set, as defined in Section \ref{sec:photometry} and Table \ref{tab:table_1}. \\

$^b$ Star Cluster ID \# from \citet{Maschmann24}.\\

$^c$ Adopted age as discussed in Sections \ref{sec:ages}. The values with ``.1" attached  are the outliers discussed in Section \ref{sec:outliers}. 

 $^d$ The 0.814-flux in $\mu$Janskys can be used to  convert the relative flux values in this table to absolute values.
\end{table*}

\begin{table*}
\caption{Normalized Flux Values Using Convolved Aperture Photometry    for the Training Set$^a$ }
\label{tab:table_8}
 
     \centering

\begin{tabular}{rrrrrrrrrrrrrrrrrr}
  \hline
ID$^b$  & Age$^c$ & 0.275 & 0.336 & 0.435 & 0.555  & 0.814 & 2.0 &  3.0  & 3.3 & 3.6 & 7.7 & 10.0 & 11.3 & 21.0 & 0.814-flux $^d$ \\

\hline  
 
%%  1724 & 1.0 & 5.15 & 4.07 & 2.38 & 1.9 & 26.48 & 1.0 & 0.7 & 0.55 & 1.78 & 1.1 & 27.79 & 12.5 & 40.7 & 84.4 & 2.86E-12 \\
    1724 &  1.0 &  0.76 & 1.07 & 1.89 & 1.48 & 1.0 & 7.94 & 6.18 & 17.54 & 11.65 & 170.65 & 65.78 & 179.26 & 297.06 & 0.659\\
  1767 & 1.0 & 0.44 & 0.66 & 1.35 & 1.16 & 1.0 & 4.77 & 4.06 & 13.27 & 8.87 & 170.36 & 77.19 & 210.48 & 724.12 & 3.890\\
  4505 &  1.0 & 0.48 & 0.73 & 1.51 & 1.28 & 1.0 & 3.82 & 3.16 & 11.29 & 7.02 & 149.98 & 57.16 & 179.96 & 380.99 & 1.623\\
  5418 &  1.0 & 0.44 & 0.66 & 1.4 & 1.17 & 1.0 & 3.35 & 2.48 & 7.52 & 4.99 & 94.19 & 34.1 & 105.32 & 300.53 & 2.744\\
  7280 &  1.0 & 0.34 & 0.52 & 0.96 & 0.9 & 1.0 & 7.23 & 11.37 & 23.91 & 20.93 & 286.57 & 147.45 & 375.72 & 984.66 & 2.350\\
  3150 &  2.0 & 0.48 & 0.74 & 1.47 & 1.31 & 1.0 & 4.72 & 4.13 & 11.81 & 7.99 & 152.38 & 48.96 & 179.95 & 345.81 & 2.744\\
  3538 &  2.0 & 0.49 & 0.75 & 1.52 & 1.34 & 1.0 & 4.67 & 4.18 & 10.2 & 6.6 & 108.47 & 42.77 & 132.25 & 435.77 & 2.558\\
  4995 &  2.0 & -0.37 & -0.46 & -0.42 & 0.04 & 1.0 & 2.16 & 0.45 & -4.24 & -1.95 & -114.35 & -37.51 & -135.77 & -318.18 & -0.806\\
  5015 &  2.0 & 0.41 & 0.62 & 1.36 & 1.27 & 1.0 & 3.77 & 4.06 & 9.82 & 9.1 & 96.17 & 46.06 & 115.92 & 147.92 & 0.960\\
  6681 &  2.0 & 0.47 & 0.69 & 1.45 & 1.29 & 1.0 & 3.78 & 3.04 & 9.59 & 6.21 & 125.84 & 45.96 & 146.55 & 366.24 & 3.421\\
  1905 &  3.0 & 0.51 & 0.76 & 1.52 & 1.28 & 1.0 & 4.3 & 3.6 & 9.67 & 6.72 & 118.68 & 38.15 & 124.84 & 160.91 & 0.707\\
  1343 &  3.0 & 0.65 & 0.83 & 1.72 & 1.39 & 1.0 & 2.51 & 1.21 & 0.71 & 0.76 & -2.77 & -1.49 & -5.22 & 0.95 & 2.415\\
  2972 &  3.0 & 0.42 & 0.66 & 1.46 & 1.3 & 1.0 & 4.61 & 4.42 & 4.68 & 4.22 & 8.92 & 9.41 & 13.99 & 23.17 & 6.860\\
  4768 &  3.1 & 0.63 & 0.85 & 1.66 & 1.46 & 1.0 & 1.04 & -0.13 & 0.69 & -1.12 & 13.65 & 2.43 & 11.39 & 52.05 & 1.072\\
  5434 &  3.0 & 0.38 & 0.57 & 1.19 & 1.13 & 1.0 & 4.23 & 2.86 & 6.49 & 4.45 & 79.04 & 29.57 & 87.06 & 278.79 & 2.746\\
  1214 &  4.0 & 0.2 & 0.34 & 0.94 & 0.88 & 1.0 & 5.43 & 5.82 & 5.49 & 5.84 & 2.06 & 15.78 & 9.69 & 18.18 & 7.631\\
  1248 &  4.0 & 0.46 & 0.74 & 1.66 & 1.41 & 1.0 & 1.43 & 0.56 & 0.17 & 0.52 & 6.56 & 0.56 & 3.01 & 51.91 & 1.256\\
  5655 &  4.0 & 0.4 & 0.62 & 1.39 & 1.22 & 1.0 & 4.0 & 2.47 & 4.12 & 3.19 & 42.77 & 16.31 & 51.25 & 159.74 & 5.160\\
  6276 &  4.0 & 0.59 & 0.82 & 1.72 & 1.36 & 1.0 & 6.61 & 4.69 & 4.47 & 4.24 & 0.24 & 4.98 & 2.23 & 6.63 & 4.642\\
  7255 &  4.1 & 0.36 & 0.45 & 1.23 & 0.99 & 1.0 & 4.06 & 2.36 & -1.12 & 0.24 & -38.48 & -10.5 & -46.43 & 8.2 & 1.281\\
  1667 &  5.0 & 0.82 & 1.12 & 2.07 & 1.77 & 1.0 & 5.33 & 2.68 & 3.48 & 2.41 & 28.03 & 12.64 & 38.49 & 9.52 & 0.457\\
  2416 &  5.0 & 0.2 & 0.35 & 1.03 & 0.97 & 1.0 & 8.23 & 4.99 & 4.39 & 4.16 & -1.75 & 4.85 & 0.9 & 1.04 & 2.578\\
  3433 &  5.0 & 0.36 & 0.71 & 1.65 & 1.4 & 1.0 & 2.12 & 1.66 & 2.94 & 2.03 & 22.31 & 9.17 & 28.56 & 34.19 & 1.030\\
  5894 &  5.0 & 0.17 & 0.28 & 1.02 & 0.95 & 1.0 & 3.87 & 1.97 & 1.61 & 1.43 & -0.14 & 1.11 & 0.67 & -2.73 & 1.899\\
  6895 &  5.0 & 0.65 & 0.98 & 2.18 & 1.67 & 1.0 & -0.32 & -0.42 & -0.23 & -0.28 & 6.98 & 2.06 & 9.35 & 4.59 & 0.977\\
  1187 &  10.0 & 0.23 & 0.3 & 0.81 & 0.78 & 1.0 & 5.72 & 2.68 & 2.39 & 2.1 & 4.18 & 1.11 & 4.62 & 3.87 & 1.999\\
  2688 &  10.0 & 0.15 & 0.23 & 0.63 & 0.64 & 1.0 & 7.95 & 4.29 & 3.66 & 3.32 & 0.41 & 1.31 & 0.36 & 2.43 & 3.369\\
  4356 &  10.0 & 0.11 & 0.18 & 0.44 & 0.43 & 1.0 & 10.16 & 5.56 & 4.9 & 4.56 & 1.6 & 3.8 & 2.65 & 1.84 & 2.542\\
  5016 &  10.0 & 0.16 & 0.23 & 0.67 & 0.65 & 1.0 & 7.67 & 3.93 & 3.26 & 2.9 & -0.6 & 0.89 & -1.71 & -0.58 & 2.436\\
  7150 &  10.0 & 0.21 & 0.35 & 1.08 & 1.07 & 1.0 & 1.53 & 0.87 & 1.12 & 0.77 & 10.12 & 4.47 & 16.59 & 6.66 & 1.906\\
  1743 &  150.0 & 0.14 & 0.22 & 0.84 & 0.86 & 1.0 & 4.81 & 2.64 & 3.48 & 2.5 & 37.29 & 12.54 & 32.84 & 30.53 & 0.851\\
  2535 &  150.0 & 0.04 & 0.11 & 0.7 & 0.81 & 1.0 & 3.41 & 1.73 & 1.14 & 1.04 & 18.13 & 5.0 & 22.8 & 1.1 & 0.449\\
  4901 &  150.0 & 0.05 & 0.13 & 0.69 & 0.81 & 1.0 & 2.17 & 0.9 & 0.49 & 0.69 & -12.48 & -4.26 & -18.57 & -5.95 & 0.724\\
  5736 &  150.0 & 0.04 & 0.11 & 0.7 & 0.89 & 1.0 & 3.65 & 1.6 & 0.99 & 0.38 & -10.97 & -6.75 & -22.42 & -1.76 & 0.772\\
  6400 &  150.0 & 0.14 & 0.26 & 0.9 & 0.92 & 1.0 & 5.87 & 3.58 & 3.19 & 2.62 & -0.77 & 3.13 & -0.94 & 3.0 & 1.695\\
  1516 &  500.0 & 0.01 & 0.05 & 0.64 & 0.77 & 1.0 & 10.98 & 6.91 & 7.27 & 7.38 & 1.23 & 2.2 & 3.0 & -1.39 & 0.632\\
  2273 &  500.0 & 0.01 & 0.05 & 0.52 & 0.68 & 1.0 & 4.17 & 2.79 & 3.12 & 3.14 & 11.13 & 7.02 & 18.98 & 2.53 & 1.395\\
  2352 &  500.1 & 0.01 & 0.06 & 0.49 & 0.77 & 1.0 & 5.43 & 4.24 & 10.9 & 7.56 & 120.3 & 44.74 & 146.18 & 269.93 & 1.412\\
  2984 &  500.0 & 0.01 & 0.05 & 0.41 & 0.71 & 1.0 & 6.67 & 3.97 & 4.02 & 3.63 & -0.67 & 2.87 & -1.01 & -4.23 & 0.645\\
  6340 &  500.0 & 0.05 & 0.09 & 0.56 & 0.79 & 1.0 & 8.28 & 4.23 & 4.02 & 3.48 & 6.66 & 1.85 & 10.99 & -1.15 & 0.718\\

  \hline

 \hline
 
\end{tabular}

 \raggedright

\bigskip

$^a$ Convolved (using images convolved to the F2100W scale of 0.67$^{\prime}{\prime}$) photometry from \citet{hassani23} for clusters in the training set, as defined in Section \ref{sec:photometry} and Table \ref{tab:table_1}.  \\
$^b$ Star Cluster ID \# from \citet{Maschmann24}.\\
$^c$ Adopted age as discussed in Sections \ref{sec:ages}. The values with ``.1" attached  are the outliers discussed in Section \ref{sec:outliers}. \\

$^d$ The 0.814-flux in $\mu$Janskys can be used to  convert the relative flux values in this table to absolute values.
 
\end{table*}

\end{document}